\documentclass[english,aps, prl, twocolumn, showpacs]{revtex4-1}
\usepackage[T1]{fontenc}
\usepackage[latin9]{inputenc}
\usepackage{color}
\usepackage{babel}
\usepackage{verbatim}
\usepackage{amsmath}
\usepackage{amssymb}
\usepackage{graphicx}
\usepackage{esint}
\usepackage[unicode=true,pdfusetitle,
 bookmarks=false,
 breaklinks=false,pdfborder={0 0 1},backref=false,colorlinks=true]
 {hyperref}

\makeatletter

\@ifundefined{textcolor}{}
{%
 \definecolor{BLACK}{gray}{0}
 \definecolor{WHITE}{gray}{1}
 \definecolor{RED}{rgb}{1,0,0}
 \definecolor{GREEN}{rgb}{0,1,0}
 \definecolor{BLUE}{rgb}{0,0,1}
 \definecolor{CYAN}{cmyk}{1,0,0,0}
 \definecolor{MAGENTA}{cmyk}{0,1,0,0}
 \definecolor{YELLOW}{cmyk}{0,0,1,0}
}

\makeatother

\begin{document}

\title{Quantum Channel Construction with Circuit Quantum Electrodynamics}

\author{Chao~Shen}

\affiliation{Department of Applied Physics and Physics, Yale University, New Haven,
Connecticut 06511, USA}

\affiliation{Yale Quantum Institute, Yale University, New Haven, Connecticut 06520,
USA}

\author{Kyungjoo~Noh}

\affiliation{Department of Applied Physics and Physics, Yale University, New Haven,
Connecticut 06511, USA}

\affiliation{Yale Quantum Institute, Yale University, New Haven, Connecticut 06520,
USA}

\author{Victor~V.~Albert}

\affiliation{Department of Applied Physics and Physics, Yale University, New Haven,
Connecticut 06511, USA}

\affiliation{Yale Quantum Institute, Yale University, New Haven, Connecticut 06520,
USA}

\author{Stefan~Krastanov}

\affiliation{Department of Applied Physics and Physics, Yale University, New Haven,
Connecticut 06511, USA}

\affiliation{Yale Quantum Institute, Yale University, New Haven, Connecticut 06520,
USA}

\author{M.~H.~Devoret}

\affiliation{Department of Applied Physics and Physics, Yale University, New Haven,
Connecticut 06511, USA}

\affiliation{Yale Quantum Institute, Yale University, New Haven, Connecticut 06520,
USA}

\author{R.~J.~Schoelkopf}

\affiliation{Department of Applied Physics and Physics, Yale University, New Haven,
Connecticut 06511, USA}

\affiliation{Yale Quantum Institute, Yale University, New Haven, Connecticut 06520,
USA}

\author{S.~M.~Girvin}

\affiliation{Department of Applied Physics and Physics, Yale University, New Haven,
Connecticut 06511, USA}

\affiliation{Yale Quantum Institute, Yale University, New Haven, Connecticut 06520,
USA}

\author{Liang~Jiang}

\affiliation{Department of Applied Physics and Physics, Yale University, New Haven,
Connecticut 06511, USA}

\affiliation{Yale Quantum Institute, Yale University, New Haven, Connecticut 06520,
USA}
\begin{abstract}
\textcolor{black}{Quantum channels can describe all transformations
allowed by quantum mechanics. We provide an explicit universal protocol
to construct all possible quantum channels, using a single qubit ancilla
with quantum non-demolition readout and adaptive control. }Our construction
is efficient in both physical resources and circuit depth, and can
be demonstrated using superconducting circuits and various other physical
platforms. There are many applications of quantum channel construction,
including system stabilization and quantum error correction, Markovian
and exotic channel simulation, implementation of generalized quantum
measurements and more general quantum instruments. Efficient construction
of arbitrary quantum channels \textcolor{black}{opens up exciting
new possibilities for quantum control, quantum sensing and information
processing tasks.}
\end{abstract}
\maketitle
\global\long\def\bra#1{\left<#1\right|}
\global\long\def\ket#1{\left|#1\right>}
\global\long\def\trace#1{\text{Tr}#1}

\section{Introduction}

\textcolor{black}{Quantum channels or quantum operations, more formally
known as completely positive and trace preserving (CPTP) maps between
density operators \cite{Nielsen_Chuang,Watrous_lec_notes,Wilde_book},
give the most general description of quantum dynamics. For closed
quantum systems, unitary evolution is sufficient to describe the dynamics.
For open quantum systems, however, the interaction between the system
and environment leads to non-unitary evolution of the system (e.g.,
dissipation), which requires CPTP maps for full characterization.
Besides describing open system dynamics, the system dissipation can
further be engineered to protect encoded quantum information from
undesired decoherence processes }\textcolor{blue}{\cite{Diehl_nat_phy_2008,Verstraete_nat_phy_2009,Blatt_nat_phy_2010,Blatt_nature_2011,Polzik_PRL_2011,Morigi_ion_chain_PRL}}\textcolor{black}{.
Hence, it is important to systematically extend quantum control techniques
from closed to open quantum systems. }

\textcolor{black}{Theoretically, universal Lindb}ladian dynamics constructions
have been investigated \cite{markovian_qubit_pra,markovian_d_pra,zanardi_lindbladian_sim},
which can be used for stabilization of target quantum states \cite{Diehl_nat_phy_2008},
protection \textcolor{black}{of information encoded in subspaces \cite{Mirrahimi_NJP},
or even quantum information processing \cite{Diehl_nat_phy_2011,Albert_PRL_2016,Albert_Lindbladian_2016}}.
Experimentally, dissipative quantum control has been demonstrated
using various physical platforms\cite{Blatt_nat_phy_2010,Blatt_nature_2011,Polzik_PRL_2011,Guo_nat_phy_2011,Saffman_PRL_2013,Wineland_dissi_two_qubit_2013,Leghta_Science}.
Besides Lindbladian dynamics, CPTP maps also include exotic indivisible
channels that \textit{cannot} be expressed as Lindbladian channels
\cite{divisibility}. Hence, \textcolor{black}{use of Lindbladian
dynamics is insufficient to construct all CPTP maps, which require
more general techniques. }

\textcolor{black}{The textbook approach to construct all CPTP maps
for a $d$-dimensional system (with $d=2^{m}$ for a system consisting
of $m$ qubits) requires a $d^{2}$-dimensional ancilla and one round
of $SU(d^{3})$ joint unitary operation (Stinespring dilation, see
\cite{Nielsen_Chuang}). One recent work suggests that using a $d$-dimensional
ancilla and a probabilistic $SU(d^{2})$ joint unitary operation might
be sufficient for all CPTP maps, based on a mathematical conjecture
\cite{Wang_Sanders_NJP_2015}. More interestingly, the ancilla dimension
can be dramatically reduced to 2 for arbitrary system dimension $d$
\cite{lloyd_viola_pra}, if we introduce adaptive control }%
\footnote{It is sometimes called feedback control \cite{lloyd_viola_pra}.%
}\textcolor{black}{{} based on quantum non-demolition (QND) readout of
the ancilla which conditions a sequence of $SU(2d)$ unitary operations.
Besides CPTP maps, the adaptive approach can be used for generalized
quantum measurement, called Positive-Operator Valued Measure (POVM)
\cite{lloyd_viola_pra}. As detailed in Ref. \cite{binary_tree_POVM},
an explicit binary tree construction has been provided to implement
any given POVM. To achieve ultimate control of open quantum systems,
it is crucial to extend the construction to general CPTP maps.}

\textcolor{black}{}

\begin{figure}
\centering{}\includegraphics{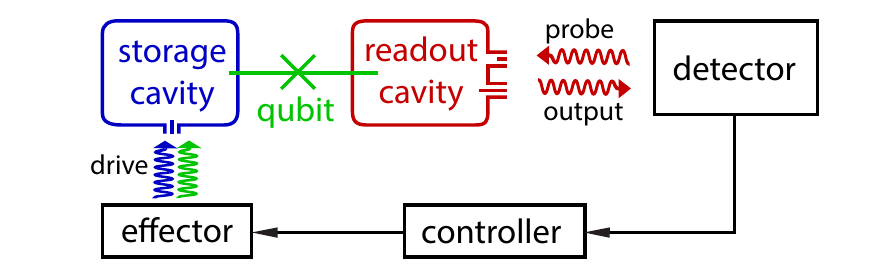}\protect\caption{(Color online) Schematic setup of a circuit QED system used for constructing
an arbitrary quantum channel. \label{fig:setup_schematic}}
\end{figure}

\textcolor{black}{In this paper, we concretize the idea developed
in \cite{lloyd_viola_pra,binary_tree_POVM} and propose a general
protocol for implementing arbitrary CPTP maps, featuring minimal physical
resources (a single ancilla qubit) and low circuit depth (logarithmic
with the system dimension). We provide an explicit proposal to implement
such a tree-like series using a minimal and currently feasible set
of operations from circuit quantum electrodynamics (cQED) \cite{Ofek_nature_2016,Wang_Science_2016,Blumoff_Chou_2016,Reinier_2016},
with the setup shown in Fig. \ref{fig:setup_schematic}.} Furthermore,
using concrete examples, we argue that the capability to efficiently
construct arbitrary CPTP maps can lead to exciting new possibilities
in the field of quantum control and quantum information processing
in general. 

The goal of this inves\textcolor{black}{tigation is to expand the
quantum control toolbox to efficiently implement all CPTP maps. In
contrast to investigations of analog/digital quantum simulators of
certain complex quantum dynamics \cite{Jaksch_PhysRevLett.81.3108,Greiner_nature2002,Buluta108,Lloyd_science_1996,Blatt_nature_2011,Weimer_nat_phy_2010,Lanyon_science2011,Wallraff_PRX_2015,Martinis_nat_comm_2015},
we focus on the efficient implementation of CPTP maps for various
quantum control tasks, including state stabilization, information
processing, quantum error correction, etc. }

This paper is organized as follows. First, we review the basic notation
of CPTP maps using the Kraus representation in Section \ref{sec:Kraus-Representation}.
We then provide an explicit protocol that can implement arbitary CPTP
maps using an ancilla qubit with QND readout and adaptive control,
and describe its implementation with cQED in Section \ref{sec:Universal-Channel-Construction}.
In Section \ref{sec:Application-Examples}, we illustrate potential
applications of such constructed CPTP maps. In Section \ref{sec:Discussion},
we discuss further extensions and various imperfections. Finally,
we conclude the paper in Section \ref{sec:Conclusion}.

\section{Kraus Representation\label{sec:Kraus-Representation}}

Mathematically, we use the Kraus representation for CPTP maps, 
\begin{equation}
\mathcal{T}(\rho)=\sum_{i=1}^{N}K_{i}\rho K_{i}^{\dagger},
\end{equation}
 which are trace-preserving as ensured by the condition\cite{Choi}
\begin{equation}
\sum_{i=1}^{N}K_{i}^{\dagger}K_{i}=\mathbb{I}.\label{eq:K-constraint}
\end{equation}
The Kraus operators $K_{i}$ do not have to be unitary or Hermitian.
They can even be non-square matrices, if the input and output Hilbert
spaces have different dimensions. By padding with zeros, we can always
make them square matrices that describe a dimension-preserving channel
for a system with dimension $d$. The Kraus representation is not
unique, because for any $N\times N$ unitary matrix $U$, the set
of new Kraus operators $F_{i}=\sum_{j}U_{ij}K_{j}$ characterizes
the same CPTP map. 

To efficiently construct a CPTP map, it is convenient to work with
the Kraus representation with the minimum number of Kraus operators,
called the \textit{Kraus rank} of the CPTP map. Since there are at
most $d^{2}$ linearly independent operators for a Hilbert space of
dimension $d$, the Kraus rank is no larger than $d^{2}$ (for a rigorous
treatment see \cite{Choi}). There are efficient procedures to convert
different representations of a channel to the minimal Kraus representation
\cite{Choi,Watrous_lec_notes,Wilde_book}. For example, we may convert
the Kraus representation into the Choi matrix (a $d^{2}\times d^{2}$
Hermitian matrix) and from there obtain the minimal Kraus representation
\cite{Choi}. The second approach is to calculate the overlap matrix
$C_{ij}=\mathrm{Tr}(K_{i}K_{j}^{\dagger})$ and then diagonalize it,
$C=V^{\dagger}DV$ \cite{Nielsen_Chuang}. The new Kraus operators,
$\tilde{K}_{i}=\sum_{j}V_{ij}K_{j}$, will be the most economic representation
with some of them being zero matrices if the original representation
is redundant. For cases with the CPTP map provided in other representations
(e.g., super-operator matrix representation, Jamiolkowski/Choi matrix
representation), we can also perform a well-defined routine to bring
them into the minimal Kraus representation (as detailed in Appendix
\ref{Appendix: representation channels}).%

\section{Universal Construction of Quantum Channels\label{sec:Universal-Channel-Construction}}

\begin{figure*}
\begin{centering}
\includegraphics{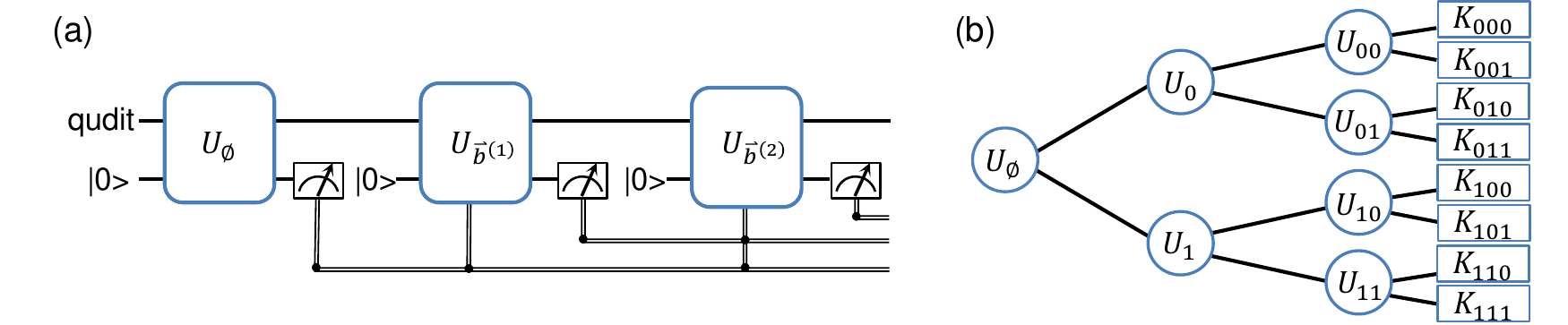}
\par\end{centering}

\protect\caption{(Color online) \textbf{(a)} Quantum circuit for arbitrary channel
construction. The dimension of the system $d$ can be arbitrary and
the circuit depth depends only on the Kraus rank of the target channel.
\textbf{(b)} Binary tree representation with depth $L=3$. The Kraus
operators $K_{b^{\left(L\right)}}$ are associated with the leaves
of the binary tree, $b^{\left(L\right)}\in\left\{ 0,1\right\} ^{L}$.
The system-ancilla joint unitary to apply in $l$-th round $U_{b^{\left(l\right)}}$
depends on the previous ancilla readout record $b^{\left(l\right)}=(b_{1}b_{2}\cdots b_{l})\in\left\{ 0,1\right\} ^{l}$
associated with a node of the binary tree. For any given channel,
all these unitaries can be explicited constructed and efficiently
implemented. \label{fig:Quantum-circuit-for} }
\end{figure*}

As first pointed out by Lloyd and Viola \cite{lloyd_viola_pra}, repeated
application of Kraus rank-2 channels in an adaptive fashion is in
principle sufficient to construct arbitrary open-system dynamics.
Andersson and Oi provided a scheme for a binary-tree construction
to explicitly implement an arbitrary POVM \cite{binary_tree_POVM}.
We extend the binary-tree scheme to a more general protocol for arbitrary
CPTP maps. The procedure to construct a CPTP map with Kraus rank
$N$ is associated with a binary tree of depth $L=\left\lceil \log_{2}N\right\rceil $,
as shown in Fig. \ref{fig:Quantum-circuit-for}. In the following,
we first consider the simple case with $L=1$, corresponding to the
CPTP maps with Kraus rank $N\le2$. Then, we provide an explicit construction
for general CPTP maps. After that, we outline how to physically implement
the circuits using cQED as a promising physical platform. 

\subsection{Quantum Channels with Kraus Rank 2 }

Given a single use of the ancilla qubit, we can construct any rank-2
CPTP map, characterized by Kraus operators $\left\{ K_{1},K_{2}\right\} $.
The procedure consists of the following: (1) initialize the ancilla
qubit in $\ket 0$, (2) perform a joint unitary operation $U\in SU(2d)$,
and (3) discard (``trace over'') the ancilla qubit. Since this procedure
has only one round of operation, there is no need for adaptive control
and thus we can simply discard the ancilla without any measurement.

The $2d\times2d$ matrix of unitary operation has the following block
matrix form %
\footnote{We choose the ordering of tensor product to be $\text{ancilla}\otimes\text{system}$.%
},
\begin{equation}
U=\left(\begin{array}{cc}
\bra 0U\ket 0 & *\\
\bra 1U\ket 0 & *
\end{array}\right),\label{eq:unitary_rank2_channel}
\end{equation}
where the $d\times d$ submatrices are $\bra 0U\ket 0=K_{0}$, $\bra 1U\ket 0=K_{1}$,
and ``{*}'' denotes irrelevant submatrices (as long as $U$ is unitary).
The trace preserving requirement, $K_{0}^{\dagger}K_{0}+K_{1}^{\dagger}K_{1}=\mathbb{I}$,
ensures that the condition $\sum_{b=0,1}\left(\bra bU\ket 0\right)^{\dagger}\bra bU\ket 0=\mathbb{I}_{d\times d}$
is fulfilled for unitary $U$. After discarding the ancilla qubit,
the procedure achieves the CPTP map,
\[
\mathcal{T}_{U}(\rho)=K_{0}\rho K_{0}^{\dagger}+K_{1}\rho K_{1}^{\dagger}.
\]
Therefore, any channel with Kraus rank 2 can be simulated with a
single use of the ancilla qubit %
\footnote{We remark that Eq. (\ref{eq:unitary_rank2_channel}) indicates that
only the left half of the unitary matrix matters and we do not really
require the capability to implement an arbitrary unitary evolution
on the combined system to simulate all rank-2 channels. We will have
more discussion on this in Sec. \ref{sub:Physical-Implementation-Example}.%
}.

If we measure the ancilla qubit instead of discarding it, we can in
principle obtain the ``which trajectory'' information. More specifically,
the system state becomes $\left(\bra 0U\ket 0\right)\rho\left(\bra 0U^{\dagger}\ket 0\right)$
(unnormalized) if we find the ancilla in $\ket 0$, and it becomes
$\left(\bra 1U\ket 0\right)\rho\left(\bra 0U^{\dagger}\ket 1\right)$
if we find the ancilla in $\ket 1$. We may use the ``which trajectory''
information to determine later operations, and thus construct more
complicated CPTP maps with higher Kraus rank.

\subsection{Quantum Channels with Higher Kraus Rank}

To implement a CPTP map with Kraus rank $N$, we need a quantum circuit
with $L=\left\lceil \log_{2}N\right\rceil $ rounds of operations.
Each round consists of (1) initialization of the ancilla qubit, (2)
joint unitary gate over the system and ancilla (conditional on the
measurement outcomes from previous rounds), (3) QND readout of the
ancilla, and (4) storage of the classical measurement outcome for
later use. For a quantum circuit consisting of $L$ rounds of operations
with adaptive control (based on binary outcomes), there are $2^{L}-1$
possible intermediate unitary gates (associated with $2^{L}-1$ nodes
of a depth-$L$ binary tree) and $2^{L}$ possible trajectories (associated
with the $2^{L}$ leaves of the binary tree).

As illustrated in Fig.~\ref{fig:Quantum-circuit-for}, we denote
the $l$-th round unitary gate as $U_{b^{\left(l\right)}}$, associated
with the node of the binary tree, $b^{\left(l\right)}=(b_{1}b_{2}\cdots b_{l})\in\left\{ 0,1\right\} ^{l}$
with $l=0,\cdots,L-1$. (For $L=1$, there is only one unitary gate
for $b^{\left(0\right)}=\varnothing$, which is $U_{b^{\left(0\right)}=\varnothing}$
as given in Eq. (\ref{eq:unitary_rank2_channel}).) Generally, the
unitary gate, $U_{b^{\left(l\right)}}$, has the following block matrix
form
\begin{equation}
U_{b^{\left(l\right)}}=\left(\begin{array}{cc}
\bra 0U_{b^{\left(l\right)}}\ket 0 & *\\
\bra 1U_{b^{\left(l\right)}}\ket 0 & *
\end{array}\right),
\end{equation}
where ``{*}'' again denote irrelevant submatrices (as long as $U_{b^{\left(l\right)}}$
is unitary). Since the ancilla always starts in $\ket 0$, it is sufficient
to specify the $d\times d$ submatrices $\bra{b_{l+1}}U_{b^{\left(l\right)}}\ket 0$
acting on the system, with the projectively measured ancilla state
$\ket{b_{l+1}}$ for $b_{l+1}=0,1$. Associated with the leaves of
the binary tree, $b^{\left(L\right)}\in\left\{ 0,1\right\} ^{L}$,
are Kraus operators labeled in binary notation, 
\begin{equation}
K_{b^{\left(L\right)}}=K_{i},
\end{equation}
with $i=\left(b_{1}b_{2}\cdots b_{L}\right)_{2}+1$ and $K_{i>N}=0$.
The singular value decomposition of each Kraus operator is $K_{b^{\left(L\right)}}=W_{b^{\left(L\right)}}D_{b^{\left(L\right)}}V_{b^{\left(L\right)}}^{\dagger}$.

We now provide an explicit construction for $\bra{b_{l+1}}U_{b^{\left(l\right)}}\ket 0$.
First, for each node $b^{\left(l\right)}$ with $l=1,\cdots,L-1$,
we may diagonalize the non-negative Hermitian matrix (which is associated
with the summation over all the leaves in the branch starting from
$b^{\left(l\right)}$) 
\begin{equation}
\sum_{b_{l+1},\cdots,b_{L}}K_{b^{\left(L\right)}}^{\dagger}K_{b^{\left(L\right)}}=V_{b^{\left(l\right)}}D_{b^{\left(l\right)}}^{2}V_{b^{\left(l\right)}}^{\dagger}\equiv M_{b^{\left(l\right)}}^{2},
\end{equation}
with unitary matrix $V_{b^{\left(l\right)}}$, diagonal matrix $D_{b^{\left(l\right)}}$
consisting of non-negative diagonal elements, and Hermitian matrix
$M_{b^{\left(l\right)}}=V_{b^{\left(l\right)}}D_{b^{\left(l\right)}}V_{b^{\left(l\right)}}^{\dagger}$.
For notational convenience, we introduce $P_{b^{\left(l\right)}}$
as the support projection matrix of $D_{b^{\left(l\right)}}$, with
elements 
\begin{equation}
\left(P_{b^{\left(l\right)}}\right)_{j,k}=\mathrm{sign}\left[\left(D_{b^{\left(l\right)}}\right)_{j,k}\right],
\end{equation}
where $\text{sign(0)}\equiv0$, so that $P_{b^{\left(l\right)}}^{2}=P_{b^{\left(l\right)}}$
and $P_{b^{\left(l\right)}}D_{b^{\left(l\right)}}=D_{b^{\left(l\right)}}P_{b^{\left(l\right)}}=D_{b^{\left(l\right)}}$.
The orthogonal projection is $P_{b^{\left(l\right)}}^{\perp}=\mathbb{I}-P_{b^{\left(l\right)}}$
and we also define the related projection $Q_{b^{\left(l\right)}}\equiv V_{b^{\left(l\right)}}P_{b^{\left(l\right)}}^{\perp}V_{b^{\left(l\right)}}^{\dagger}$.
In addition, we define 
\begin{equation}
\left(D_{b^{\left(l\right)}}^{-1}\right)_{j,k}=\begin{cases}
1/\left(D_{b^{\left(l\right)}}\right)_{j,k} & \mbox{if \ensuremath{\left(D_{b^{\left(l\right)}}\right)\neq0}}\\
0 & \mbox{otherwise.}
\end{cases}
\end{equation}
and denote the Moore-Penrose pseudo-inverse of $M_{b^{\left(l\right)}}$
as $M_{b^{\left(l\right)}}^{+}=V_{b^{\left(l\right)}}D_{b^{\left(l\right)}}^{-1}V_{b^{\left(l\right)}}^{\dagger}$.
For $l=0$, we fix $V_{b^{\left(0\right)}}=D_{b^{\left(0\right)}}=D_{b^{\left(0\right)}}^{-1}=P_{b^{\left(0\right)}}=\mathbb{I}$
and $P_{b^{\left(0\right)}}^{\perp}=0$.

Finally, we have the explicit expression for the relevant submatrices
of the unitary matrix 
\begin{align}
\bra{b_{l+1}}U_{b^{\left(l\right)}}\ket 0 & =M_{b^{\left(l+1\right)}}M_{b^{\left(l\right)}}^{+}+\frac{1}{\sqrt{2}}Q_{b^{\left(l\right)}}
\end{align}
with $b^{\left(l+1\right)}=\left(b^{\left(l\right)},b_{l+1}\right)$
for $l=0,\cdots,L-2$, and
\begin{align}
\bra{b_{l+1}}U_{b^{\left(l\right)}}\ket 0 & =K_{b^{\left(l+1\right)}}M_{b^{\left(l\right)}}^{+}+\frac{1}{\sqrt{2}}W_{b^{\left(l+1\right)}}V_{b^{\left(l+1\right)}}^{\dagger}Q_{b^{\left(l\right)}}\label{eq:Block4LastRound}
\end{align}
for $l=L-1$. Since the isometric condition $\sum_{b_{l+1}=0,1}\left(\bra{b_{l+1}}U_{b^{\left(l\right)}}\ket 0\right)^{\dagger}\bra{b_{l+1}}U_{b^{\left(l\right)}}\ket 0=\mathbb{I}_{d\times d}$
is fulfilled (as proven in Appendix \ref{sec:Appendix:-Proof of solution}),
we can complete the unitary matrix $U_{b^{\left(l\right)}}$ with
appropriate submatrices $\bra{b_{l+1}}U_{b^{\left(l\right)}}\ket 1$. 

For $L=1$, we use Eq. (\ref{eq:Block4LastRound}) for $l=0$ and
obtain $\bra{b_{1}}U_{b^{\left(0\right)}}\ket 0=K_{b^{\left(1\right)}}=\begin{cases}
K_{1} & \mbox{for \ensuremath{b_{1}=0}}\\
K_{2} & \mbox{for \ensuremath{b_{1}=1}}
\end{cases}$, which is consistent with the earlier construction for Kraus rank
2 channels. 

With the above explicit construction of arbitrary CPTP maps, we will
investigate the physical implementation with circuit QED.

\subsection{Physical Implementation with Circuit QED \label{sub:Physical-Implementation-Example}}

The above channel construction scheme relies on three key components:
(1) ability to apply a certain class of unitary gates (recall that
we engineer only the left half of the unitary) on the system and ancilla
combined system; (2) QND readout of the ancilla qubit; (3) adaptive
control of all unitary gates based on earlier rounds of QND measurement
outcomes. Although there are a total of $(2^{n}-1)$ unitaries potentially
to be applied, they can all be pre-calculated and one only needs to
decide which one to perform in real time based on the measurement
record. In principle any quantum system that meets these three requirements
can be used to implement our scheme. In the following, we focus on
a circuit QED system with a transmon qubit dispersively coupled to
a microwave cavity with Hamiltonian \cite{Schuster07} 
\[
\hat{H}_{0}=\omega_{c}\hat{a}^{\dagger}\hat{a}+\omega_{q}\ket e\bra e-\chi a^{\dagger}a\ket e\bra e,
\]
where $\omega_{c}$ and $\omega_{q}$ are the cavity and qubit transition
frequency respectively, $\hat{a}$ is the the annihilation operator
of a cavity excitation, $\chi$ is the dispersive shift parameter
and $\ket e\bra e$ is the qubit excited \textcolor{black}{state projection.
This is a promising platform to implement the channel construction
scheme because the dispersive shift $\chi$ can be three orders of
magnitude larger than the dissipation of the qubit and the cavity,
allowing universal unitary control of the system }\cite{Heeres_PRL,Krastanov_PRA}. 

\begin{figure}
\centering{}\includegraphics{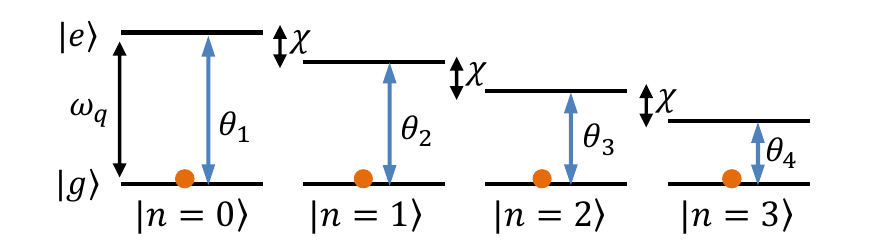}\protect\caption{\textcolor{black}{(Color online) Level diagram for the dispersively
coupled qubit-cavity system. It is straightforward to implement $U_{\mathrm{ent}}$
for such a system by driving two level transitions that are spectrally
separated. }Here $g$/$e$ denote the ancilla qubit states ($0/1$
logical states) and $n$ denotes the photon number state.\textcolor{black}{\label{fig: level diagram}}}
\end{figure}

\begin{figure*}
\centering{}\includegraphics{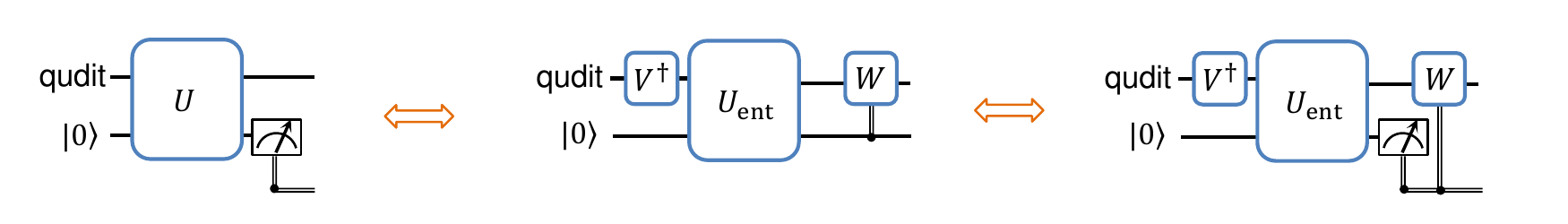}\protect\caption{\textcolor{black}{(Color online)} For circuit QED systems, the $2d$-dimensional
unitary used to generate an arbitrary Kraus rank-2 channel can be
eventually simplified to unitaries acting on the system alone and
an entangling operation $U_{ent}$ {[}Eq. (\ref{eq:Uent}){]}, which
is a series of independent two-level transitions between $\protect\ket{g,\, n}$
and $\protect\ket{e,\, n}$, where $g$/$e$ denote the ancilla qubit
states ($0/1$ logical states) and $n$ denotes the photon number
state. \label{fig: U2d_decomposition}}
\end{figure*}

The Fock states of a cavity mode can be used to encode a $d$-dimensional
system and the qubit can be used as the ancilla. Universal unitary
control on the $d$-level system has been proposed in Ref. \cite{Krastanov_PRA}
and demonstrated experimentally in Refs. \cite{Heeres_PRL,Reinier_2016}.
\textcolor{black}{The strong dispersive coupling of the cavity and
qubit enables selective driving of transitions between $\ket{g,\, n}$
and $\ket{e,\, n}$ for different excitation numbers $n$, which can
implement the following entangling unitary gate
\begin{eqnarray}
 &  & U_{\mathrm{ent}}(\theta_{i})\nonumber \\
 & = & \left(\begin{array}{c|c}
S_{0} & -S_{1}\\
\hline S_{1} & S_{0}
\end{array}\right)\nonumber \\
 & = & \left(\begin{array}{ccc|ccc}
\cos\frac{\theta_{1}}{2} &  &  & -\sin\frac{\theta_{1}}{2}\\
 & \ddots &  &  & \ddots\\
 &  & \cos\frac{\theta_{d}}{2} &  &  & -\sin\frac{\theta_{d}}{2}\\
\hline \sin\frac{\theta_{1}}{2} &  &  & \cos\frac{\theta_{1}}{2}\\
 & \ddots &  &  & \ddots\\
 &  & \sin\frac{\theta_{d}}{2} &  &  & \cos\frac{\theta_{d}}{2}
\end{array}\right)\nonumber \\
 & = & \prod_{n=0}^{d-1}\exp(-iY_{n}\theta_{n}/2),\label{eq:Uent}
\end{eqnarray}
where $Y_{n}\equiv-i\ket{g,\, n}\bra{e,\, n}+h.c.$ is the Pauli-$Y$
operator for the two-dimensional subspace associated with $n$ excitations
(see Fig.}\textcolor{red}{{} }\textcolor{black}{\ref{fig: level diagram}).
This entangling gate gives a channel described by Kraus operators
$\{S_{0},\, S_{1}\}$. If we precede $U_{\mathrm{ent}}$ with a unitary
$V^{\dagger}$ acting on the system alone and perform an adaptive
unitary on the system after $U_{\mathrm{ent}}$ depending on the ancilla
measurement $W_{0}$ or $W_{1}$, we end up with the unitary
\begin{eqnarray*}
U_{\mathrm{ent}}' & = & \left(\begin{array}{cc}
W_{0} & 0\\
0 & W_{1}
\end{array}\right)\left(\begin{array}{c|c}
S_{0} & -S_{1}\\
\hline S_{1} & S_{0}
\end{array}\right)\left(\begin{array}{cc}
V^{\dagger} & 0\\
0 & V^{\dagger}
\end{array}\right)\\
 & = & \left(\begin{array}{cc}
W_{0}S_{0}V^{\dagger} & *\\
W_{1}S_{1}V^{\dagger} & *
\end{array}\right).
\end{eqnarray*}
Remarkably, this construction is already sufficient to perfectly match
the relevant two submatrices of the desired unitary 
\[
U=\left(\begin{array}{cc}
\bra 0U\ket 0 & *\\
\bra 1U\ket 0 & *
\end{array}\right),
\]
}with $\bra 0U\ket 0=W_{0}S_{0}V^{\dagger}$ and $\bra 1U\ket 0=W_{1}S_{1}V^{\dagger}$\textcolor{black}{.
To implement the quantum circuit in Fig. \ref{fig:Quantum-circuit-for}(a),
we may explicitly identify the $W_{0/1}$, $S_{0/1}$, and $V$ matrices
for unitary operations at different rounds $U=U_{b^{\left(l\right)}}$.}

To justify the above claim, we provide an explicit design of $U_{\mathrm{ent}}'$
to perfectly match the left two submatrices of $U_{b^{\left(l\right)}}$
in three steps. (1) We start with singular value decompositions (SVD)
$\bra 0U\ket 0=W_{0}S_{0}V_{0}^{\dagger}$ and $\bra 1U\ket 0=W_{1}S_{1}V_{1}^{\dagger}$,
where we have already set the $W$'s and $S$'s to their desired values.
Now all that is left to do is to make sure that $V_{0}=V_{1}=V$.
To uniquely determine the decomposition, we require that the singular
values in $S_{0}$ are arranged in \textit{descending} order $\left(S_{0}\right)_{j,j}\ge\left(S_{0}\right)_{j+1,j+1}$,
while the singular values in $S_{1}$ are arranged in \textit{ascending}
order $\left(S_{1}\right)_{j,j}\le\left(S_{1}\right)_{j+1,j+1}$.
(2) The isometric condition $\sum_{b=0,1}\left(\bra bU\ket 0\right)^{\dagger}\bra bU\ket 0=\mathbb{I}_{d\times d}$
requires that $V_{0}^{\dagger}V_{1}S_{1}^{2}V_{1}^{\dagger}V_{0}=\mathbb{I}_{d\times d}-S_{0}^{2}.$
\textcolor{black}{Since both $S_{1}^{2}$ and $\mathbb{I}_{d\times d}-S_{0}^{2}$
are diagonal with elements in ascending order, $V_{1}^{\dagger}V_{0}$
must be the identity -- that is, $V_{0}=V_{1}=V$. Therefore, we have
obtained all the components of $U'_{\mathrm{ent}}$, which fulfills
}$\bra 0U\ket 0=W_{0}S_{0}V^{\dagger}$ and $\bra 1U\ket 0=W_{1}S_{1}V^{\dagger}$.\textcolor{black}{
A similar property was used in \cite{Korotkov_PRA_2014} to simplify
the contruction of generalized measurements of a qubit. In terms of
circuits, we decomposed the $2d$-dimensional unitaries in Fig. \ref{fig:Quantum-circuit-for}
into a series of simpler operations, as shown in Fig. \ref{fig: U2d_decomposition}. }

\textcolor{blue}{}

\section{Application Examples\label{sec:Application-Examples}}

The concept of CPTP maps encompasses all physical operations ranging
from cooling, quantum gates, measurements, to dissipative dynamics.
The capability to construct an arbitrary CPTP map offers a unified
approach to all aspects of quantum technology. To illustrate the wide
range of impact of quantum channel construction, we now investigate
some interesting applications, including quantum system initialization/stabilization,
quantum error correction, Lindbladian quantum dynamics, exotic quantum
channels, and quantum instruments.

\subsection{Initialization/Stabilization}

Almost all quantum information processing tasks require working with
a well-defined (often pure) initial state. One common approach is
to sympathetically cool the system to the ground state by coupling
to a cold bath, or optically pumping to a specific dark state, and
then performing unitary operations to bring the system to a desired
initial state. This can be slow if the system has a large relaxation
time scale. Another approach is to actively cool the system by measurement
and adaptive control. Along the line of the second approach, the channel
construction technique can be applied to discretely pump the system
from an arbitrary state into the target state $\sigma$, which can
be pure or mixed. The pumping time depends on the quantum gate and
measurement speed, instead of the natural relaxation rate. 

It is well known that the CPTP map 
\[
\rho\mapsto\mathcal{E}_{\mathrm{Init}}\left(\rho\right)=\mathrm{Tr}(\rho)\sigma
\]
stabilizes an arbitrary state $\sigma$ \cite{Watrous_lec_notes,Wilde_book}.
If the target state has diagonal representation $\sigma=\sum_{\mu}\lambda_{\mu}\ket{\psi_{\mu}}\bra{\psi_{\mu}}$,
where $\lambda_{\mu}\ge0$ and $\sum_{\mu}\lambda_{\mu}=1$, one explicit
form of Kraus operators is $\left\{ K_{i}^{\mu}=\sqrt{\lambda_{\mu}}\ket{\psi_{\mu}}\bra i\right\} $,
where $\ket i$ are a basis of the system Hilbert space \cite{Pechen_JPA_2007}.
Contrary to the conventional approaches discussed in the previous
paragraph, this dissipative map bundles the cooling and state preparation
steps and pumps an arbitrary state into state $\sigma$. Depending
on $\sigma$, entropy can be extracted from or injected into the system
by the ancilla qubit. If we run the channel construction circuit repeatedly,
state stabilization can be achieved. 

Besides pure state initialization for quantum information processing,
preparation of carefully designed mixed states may find application
in the study of foundational issues of quantum mechanics such as quantum
discord, quantum contextuality, and quantum thermodynamics \cite{Zurek_prl_2001,Vedral_JPA_2001,Kirchmair_nature_2009,Zu_PRL_2012,Horodecki_Nat_comm_2013,Brandao_PNAS_2015}.

\subsection{Quantum Error Correction}

\textcolor{black}{Besides unique steady states, there are CPTP maps
that can stabilize multiple steady states or even a subspace of steady
states, which may be used to encode useful classical or quantum information.
A practically useful application of such CPTP maps with subspaces
of steady states is quantum error correction (QEC).  Typical QEC
schemes encode quantum information in some carefully chosen logical
subspaces \cite{Gottesman97,Nielsen_Chuang} (or subsystems \cite{Bacon06}),
and use syndrome measurement and conditional recovery operations to
actively decouple the system from the environment. Despite the variety
of QEC codes and recovery schemes, the operation of any QEC recovery
can always be identified as a quantum channel. }

\textcolor{black}{For qubit-based stabilizer codes with $N_{s}$ stabilizer
generators, the recovery is a CPTP map with Kraus rank $2^{N_{s}}$
\cite{Nielsen_Chuang}. We may first use the ancilla to sequentially
measure all $N_{s}$ stabilizer generators to extract the syndrome,
and finally perform a correction unitary operation conditioned on
the syndrome pattern. Since the stabilizer generators commute with
each other, their ordering does not change the syndrome. Moreover,
the stabilizer measurement does not require conditioning on previous
measurement outcomes, because the unitary operation at the $l$-th
round is simply $U_{b^{\left(l\right)}}=U_{l}=P_{+}\otimes\hat{S}_{l}+P_{-}\otimes I$
with $\hat{S}_{l}$ for the $l$-th stabilizer and $P_{\pm}=\frac{1}{2}\left(\ket g\pm\ket e\right)\left(\bra g+\bra e\right)$,
which is independent of the previous measurement outcomes $b^{\left(l-1\right)}$.
Finally, we perform the correction unitary operation $U_{b^{\left(N_{s}\right)}}$
conditioned on the syndrome $b^{\left(N_{s}\right)}$.}

\textcolor{black}{Generally, we may consider all QEC codes that fulfill
the quantum error-correction conditions associated with a set of error
operations \cite{Knill97,Nielsen_Chuang}. For these QEC codes, we
can explicitly obtain the Kraus representation of the QEC recovery
map \cite{Knill97,Nielsen_Chuang}, which can be efficiently implemented
with our construction of quantum channels. For example, let us consider
the binomial code \cite{Girvin_PRX_2016}, which uses the larger Hilbert
space of higher excitations to correct excitation loss errors in bosonic
systems. In order to correct up to two excitation losses, the binomial
code encodes the two logical basis states as}

\textcolor{black}{
\begin{eqnarray*}
\ket{W_{\uparrow}} & \equiv & \frac{\ket 0+\sqrt{3}\ket 6}{2},\\
\ket{W_{\downarrow}} & \equiv & \frac{\sqrt{3}\ket 3+\ket 9}{2}.
\end{eqnarray*}
For small loss probability $\gamma$ for each excitation, this encoding
scheme can correct errors up to $O\left(\gamma^{2}\right)$, which
includes the following four relevant processes: identity evolution
($\hat{I}$), losing one excitation ($\hat{a}$), losing two excitations
($\hat{a}^{2}$), and back-action induced dephasing ($\hat{n}$) \cite{Girvin_PRX_2016}.
Based on the Kraus representation of the QEC recovery (with Kraus
rank 4), we can obtain the following set of unitary operations $U_{b^{\left(l\right)}}$
for the construction of the QEC recovery channel with an adaptive
quantum circuit: 
\begin{eqnarray*}
\tilde{U}_{\varnothing} & = & \left(\begin{array}{c}
\hat{P}_{3}\\
\hat{I}-\hat{P}_{3}
\end{array}\right),\\
\tilde{U}_{0} & = & \left(\begin{array}{c}
\hat{P}_{W}\\
\hat{I}-\hat{P}_{W}
\end{array}\right),\ \tilde{U}_{1}=\left(\begin{array}{c}
\hat{P}_{1}\\
\hat{I}-\hat{P}_{1}
\end{array}\right),\\
\tilde{U}_{00} & = & \left(\begin{array}{c}
\hat{I}\\
\hat{0}
\end{array}\right),\ \tilde{U}_{01}=\left(\begin{array}{c}
U_{\hat{n}}\\
\hat{0}
\end{array}\right),\\
\tilde{U}_{10} & = & \left(\begin{array}{c}
U_{\hat{a}}\\
\hat{0}
\end{array}\right),\ \tilde{U}_{11}=\left(\begin{array}{c}
U_{\hat{a}^{2}}\\
\hat{0}
\end{array}\right),
\end{eqnarray*}
where the projections are defined as $\hat{P}_{i}\equiv\sum_{k}\ket{3k+i}\bra{3k+i}$
and $\hat{P}_{W}\equiv\ket{W_{\uparrow}}\bra{W_{\uparrow}}+\ket{W_{\downarrow}}\bra{W_{\downarrow}}$,
and the unitary operators $U_{\hat{O}}$ ($\hat{O}=\hat{a},\,\hat{a}^{2},\,\hat{n}$)
transform the error states $\hat{O}\ket{W_{\sigma}}$ back to $\ket{W_{\sigma}}$
for $\sigma=\uparrow,\,\downarrow$. Explicitly, 
\[
U_{\hat{O}}=\sum_{\sigma}\ket{W_{\sigma}}\frac{\bra{W_{\sigma}}\hat{O}^{\dagger}}{\sqrt{\bra{W_{\sigma}}\hat{O}^{\dagger}\hat{O}\ket{W_{\sigma}}}}+U^{\perp},
\]
 where $U^{\perp}$ is any isometry that takes the complement of the
syndrome subspace to the complement of the logical subspace. In the
first two rounds, we perform the projective measurements to extract
the error syndrome. In the last round, we apply a correction unitary
operation to restore the logical states. Specifically, if the measurement
outcome $b^{\left(2\right)}=(0,0)$, there is no error and identify
operation $\hat{I}$ is sufficient. If $b^{\left(2\right)}=(0,1)$,
there is back-action induced dephasing error, which changes the coefficients
of Fock states so we need to correct for that with $U_{\hat{n}}$.
If $b^{\left(2\right)}=(1,1)$, there is a single excitation loss,
which can be fully corrected with $U_{\hat{a}}$. If $b^{\text{\ensuremath{\left(2\right)}}}=\left(1,0\right)$,
there are two excitation losses, which can be fully corrected with
$U_{\hat{a}^{2}}$. Repetitive application of the above QEC recovery
channel can stabilize the system in the code space spanned by $\ket{W_{\uparrow}}$
and $\ket{W_{\downarrow}}$. }

\textcolor{black}{More interestingly, beyond exact QEC codes there
are approximate QEC codes \cite{Leung97,Ng10,Beny10,Mandayam12},
which can also efficiently correct errors but only approximately fulfill
the QEC criterion. For approximate QEC codes, it is very challenging
to analytically obtain the optimal QEC recovery map, but one can use
semi-definite programming to numerically optimize the entanglement
fidelity and obtain the optimal QEC recovery map \cite{Fletcher07,FletcherThesis07,Albert16x,CPTP_SDP_2002}.
Alternatively one can use the transpose channel \cite{trans_channel_AQEC}
or quadratic recovery channels \cite{Beny10,Beny_PRA_2011,Tyson_JMP_2010}
which are known to be near-optimal. All these recovery channels can
be efficiently implemented with our general construction of CPTP maps. }

\subsection{Markovian Channels}

Recently, there has been growing interest in designing and engineering
open system dynamics for quantum information processing \cite{markovian_d_pra,markovian_qubit_pra,Verstraete_nat_phy_2009,Diehl_nat_phy_2011,Paulisch_NJP_2016},
which uses Markovian channels 
\[
\rho\rightarrow\mathcal{E}_{\mathrm{MC},t}\left(\rho\right)=\mathbb{T}\left[e^{\int_{0}^{\tau}\mathcal{L}_{t}dt}\right]\rho,
\]
where $\mathbb{T}$ stands for time ordering, and $\mathcal{L}_{t}$
is the time-dependent Lindbladian operator that has general form

\begin{eqnarray*}
\mathcal{L}_{t}(\rho) & = & -\frac{i}{\hbar}\left[H,\,\rho\right]\\
 &  & +\sum_{n,m}h_{n,m}\left[L_{n}\rho L_{m}^{\dagger}-\frac{1}{2}(\rho L_{m}^{\dagger}L_{n}+L_{m}^{\dagger}L_{n}\rho)\right],
\end{eqnarray*}
where $L_{n}$ are jump operators. Markovian channels are a special
class of CPTP maps \cite{divisibility}. In contrast to the continuous
time evolution approach \cite{markovian_qubit_pra,markovian_d_pra,zanardi_lindbladian_sim},
we construct $\mathcal{E}_{\mathrm{MC},t}=\mathcal{T}\left[e^{\int_{0}^{\tau}\mathcal{L}_{t}dt}\right]$
directly, which is advantageous in that it does not take more time
to see results for larger $\tau$ because no Trotterization or stroboscopic
control is required. We consider the following cat-pumping example
to manifest these points. 

Using a specifically engineered dissipation for a cavity mode, one
can stabilize a two-dimensional steady-state subspace spanned by the
so called cat-code \cite{Mirrahimi_NJP,Leghta_Science}. The required
dissipation can be described by the following \textit{time-independent}
Lindbladian,
\[
\mathcal{L}(\rho)=J\rho J^{\dagger}-\frac{1}{2}(J^{\dagger}J\rho+\rho J^{\dagger}J),
\]
where the jump operator $J$ is 
\[
J=\sqrt{\kappa}\prod_{i}^{n}(a-\alpha_{i}).
\]
The complex variables $\alpha_{i}$ determine the coherent state components
$\ket{\alpha_{i}}$ that span the steady-state subspace. As proposed
in \cite{Mirrahimi_NJP} and demonstrated in \cite{Leghta_Science},
the dissipation can be engineered by coupling the system mode and
another lossy mode with Hamiltonian $H=J^{\dagger}b+h.c.$ where $b$
is the annihilation operator for the lossy mode. Practically, it is
challenging to generate desired engineered dissipation that is much
stronger than the undesired dissipations (e.g., dephasing, Kerr effect,
etc). In addition, it is difficult to extract the Hamiltonian $H$
associated with higher-order nonlinearity, in order to have a higher-dimensional
steady state subspace with more coherent components. With our approach,
however, the effective rate $\kappa$ can be large and determined
by the time scale to implement the circuits, which is limited by the
duration of gates and measurements, and the delay of adaptive control.
Moreover, the construction can easily extend to the case that simultaneously
stabilizes many coherent components.

\begin{figure*}
\centering{}\includegraphics{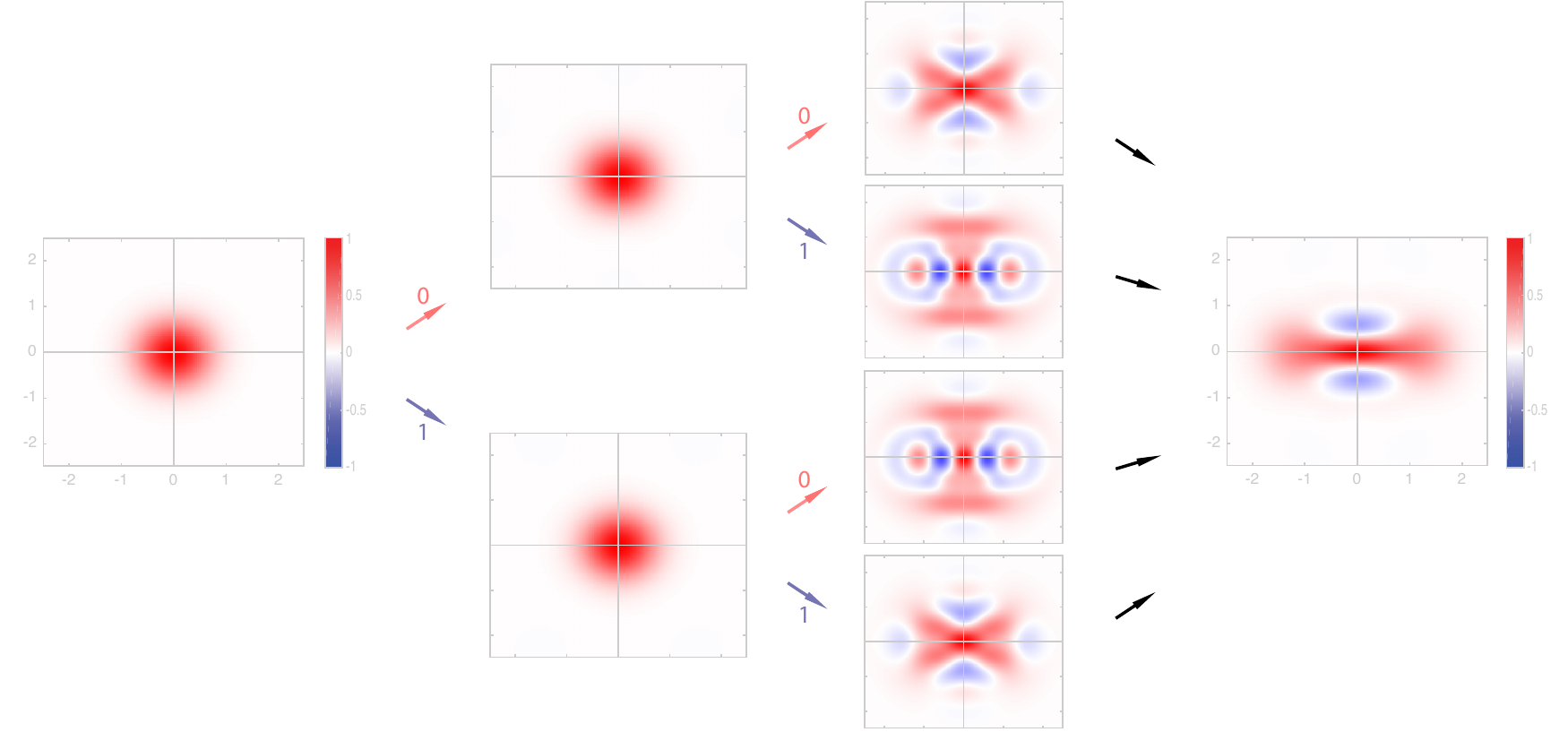}\protect\caption{(Color online) All possible trajectories for pumping a vacuum state
$\protect\ket 0$ to the subspace spanned by $\protect\ket{\pm\alpha}$
with $\alpha=1.1$. Depending on the probabilistic ancilla readout,
the system evolves along different trajectories in each run of the
circuit.\textcolor{black}{{} However, since the steady state of the
system is a pure state $\protect\ket{\psi_{\mathrm{f}}}=\left(\protect\ket{\alpha}+\protect\ket{-\alpha}\right)/\sqrt{2}$, which
cannot be decomposed as a probabilistic mixture of different states,
the final state for each trajectory is always the same pure state
$\protect\ket{\psi_{\mathrm{f}}}$. The two outcomes of the first
round are only slightly different. Two of the four outcomes of the
second round are also very similar to the others. }\label{fig:cat_pumping_example}}
\end{figure*}

\begin{figure*}
\centering{}\includegraphics{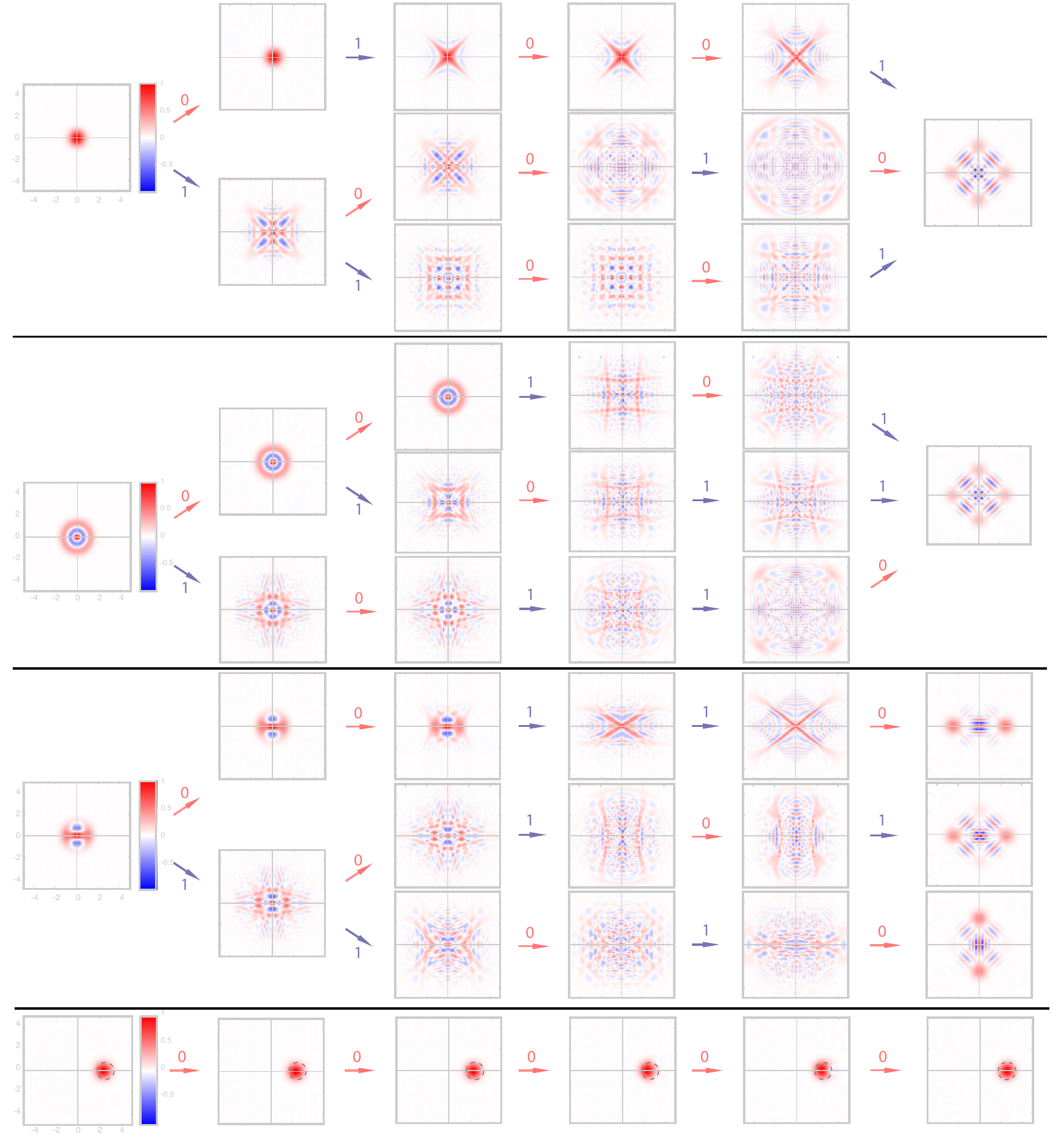}\protect\caption{(Color online) Example trajectories for 4-component cat pumping starting
with four different initial states, $\protect\ket 0,$ $\protect\ket 2$, $\left(\protect\ket 0+\protect\ket 2\right)/\sqrt{2}$
and coherent state $\protect\ket{\tilde{\alpha}=2.3}$. Here the steady
coherent components are $\protect\ket{\alpha}$, $\protect\ket{i\alpha}$,
$\protect\ket{-\alpha}$, and $\protect\ket{-i\alpha}$ with $\alpha=2.5$.
The binary number on the arrow indicates the ancilla measurement outcome.
\textcolor{black}{For the first two cases, since the steady state
is a pure state which cannot be decomposed as a probabilistic mixture
of different states, the final state for each trajectory is always
the same pure state $\protect\ket{\psi_{\mathrm{f}}}$. For the third
case, the steady state is a mixed state $\rho_{\mathrm{f}}$, so different
trajectories give different pure states. Since the ancilla measurement
results are discarded, the output state for the system is an ensemble
of the different final states, which coincides with $\rho_{\mathrm{f}}$.
The fourth case starts near the steady state subspace and is slowly
pulled into it. The trajectory shown is the dominant one which is
taken with probability higher than 0.96. Dashed circles show the position
of $\protect\ket{\alpha=2.5}$. }\label{fig:4cat pumping}}
\end{figure*}

\begin{figure*}
\centering{}\includegraphics{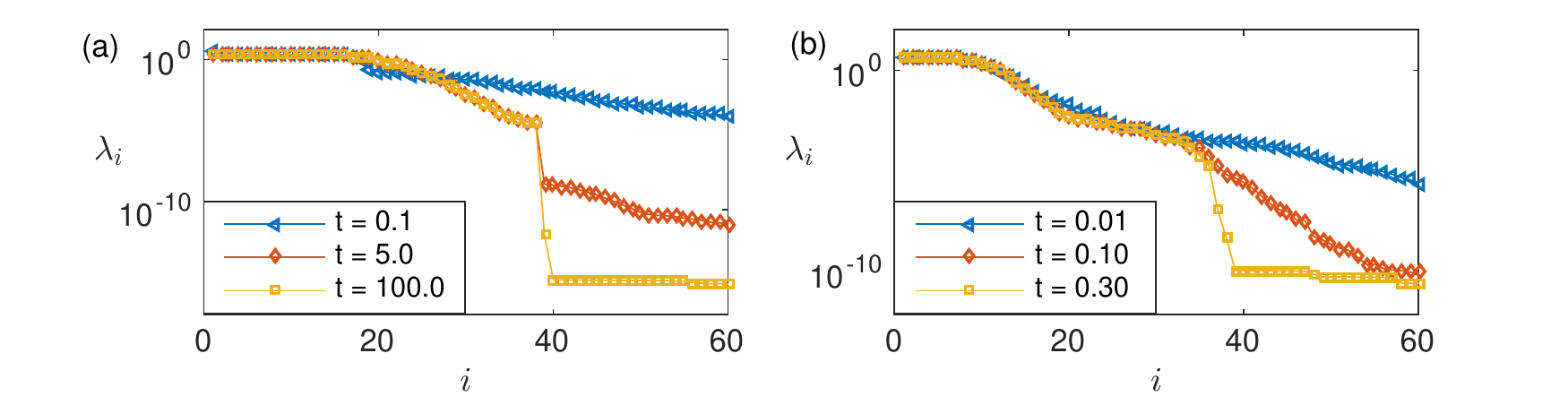}\protect\caption{(Color online) The magnitudes of the Kraus operators $\lambda_{i}\equiv\mathrm{Tr}(K_{i}^{\dagger}K_{i})$,
corresponding to $\mathcal{E}_{t}=\exp(\mathcal{L}\, t)$ for \textbf{(a)}
two-legged cat pumping and \textbf{(b)} four-legged cat pumping. Here
we set $\kappa=1$. In the long time limit, both channels have Kraus
rank approximately equal to the size of the truncated Hilbert space
$d=n_{c}+1$ where $n_{c}$ is the maximal photon number. We treat
all $\lambda_{i}$ smaller than $10^{-10}$ as 0. The figures show
results with $n_{c}=38$ but we verified that our observation remains
valid for any sufficiently large $n_{c}$. \label{fig:rank_pumping}}
\end{figure*}

With the channel construction presented here, we can now obtain Lindbladian
dynamics $\mathcal{E}_{\mathrm{MC},t}=\exp(\mathcal{L}\, t)$ for
any given $t$. \textcolor{black}{Sometimes we are interested in the
channel for $t\rightarrow\infty$ (or equivalently the strong pumping
limit $\kappa\rightarrow\infty$), $\mathcal{E}_{\mathrm{MC},\infty}$,
and it was recently shown that any more general (i.e., non-Markovian)
channel can be embedded in $\mathcal{E}_{\mathrm{MC},\infty}$ \cite{Albert_Lindbladian_2016}. }For
our approach, sending $t$ to $\infty$ does not cost us an infinite
amount of time, since the number of cycles in our construction circuit
only scales logarithmically with the Kraus rank of $\mathcal{E}_{MC,\infty}$.
In numerical calculations, the Kraus rank is not a clear-cut quantity
even when we have obtained the most economic Kraus representation.
So we define and examine the ``magnitudes'' of the Kraus operators,
$\lambda_{i}\equiv\mathrm{Tr}(K_{i}^{\dagger}K_{i})$ and remove $K_{i}$
from the description of the channel if $\lambda_{i}<10^{-10}$. Note
that $\lambda_{i}/d$ is the probability for $K_{i}$ to act on the
system when the input state is the maximally mixed state, $\rho=I/d$.
The $\lambda_{i}$ also turn out to be the eigenvalues of the Choi
matrix, see Appendix \ref{Appendix: representation channels} for
details. Numerically we found that $\mathcal{E}_{\infty}$ has lower
Kraus rank than $\mathcal{E}_{MC,t}$ with finite $t$, see Fig. \ref{fig:rank_pumping}
for two examples. In the infinite time limit, the Kraus rank scales
linearly with the dimension of the truncated Hilbert space $d=n_{c}+1$
(whre $n_{c}$ is the photon number truncation), much smaller than
the largest possible value $d^{2}$. 

\textcolor{black}{Figure \ref{fig:cat_pumping_example} and figure
\ref{fig:4cat pumping} (corresponding to $n=2$ and $n=4$ coherent
components) show trajectories }%
\footnote{\textcolor{black}{See http://qchannels.krastanov.org/ for an online
exhibition of the full trajectories. }%
}\textcolor{black}{{} of the system evolution under our constructed channel
for a large $t\sim10^{3}/\kappa$. In each run of the }simulation,\textcolor{black}{{}
the ancilla measurement results that correspond to different trajectories
are probabilistic. If the system starts in $\ket 0$, $\ket 1$, or
$\ket 2$, the correct steady state is pure. So whichever trajectory
the system follows, it ends up in the same pure state. If the system
starts in a state like $\left(\ket 0+\ket 2\right)/\sqrt{2}$, the
steady state is a mixed state, in which case different trajectories
lead to different final states. But the probabilistic mixture of all
these final states make up the expected steady state density matrix
$\rho_{\mathrm{f}}=\mathcal{E}_{t}(\rho_{\mathrm{init}})$. }

Our approach of constructing CPTP maps thus provides another promising
pathway to efficiently pump the cavity mode into the cat-code subspace
using approximately $\log_{2}(d)$ rounds of operations, each of which
consists of adaptive $SU\left(2d\right)$ unitary gates, qubit QND
measurement, and storing the measurement outcome. In the exact same
fashion, we can construct CPTP maps that manipulate the logical states
living in the code subspace, which can, e.g., implement a digital
version of holonomic gates \cite{holonomic_gates_Victor_2016}. 

\textcolor{green}{}

\subsection{Exotic Channels }

Besides Markovian channels, there are also exotic CPTP maps that cannot
be obtained from time dependent Lindbladian master equations. Hence,
these channels are not accessible in previous proposals of open system
evolution under Lindbladian master equations \textcolor{blue}{\cite{markovian_qubit_pra,markovian_d_pra,zanardi_lindbladian_sim}}.
For example, we can define the following CPTP map (called the ``partial
corner transpose'' channel) for $d$-dimensional systems \cite{divisibility}
\[
\mathcal{T}(\rho)=\frac{\rho^{T_{c}}+\mathbb{I}\,\mathrm{Tr}(\rho)}{1+d},
\]
where $\rho^{T_{c}}$ is the ``corner transposed'' density matrix
(i.e. exchanging the matrix elements $\rho_{1,d}$ and $\rho_{d,1}$
while keeping all other elements unchanged). Following Ref. \cite{divisibility},
the partial corner transpose channel has diagonal representation in
the generalized Gell-Mann basis, with identical eigenvalues $1/\left(d+1\right)$,
except for two basis elements -- the eigenvalue is $1$ for basis
element $I_{d\times d}/\sqrt{d}$, and the eigenvalue is $-1/\left(d+1\right)$
for basis element $\left(\ket d\bra 1+\ket 1\bra d\right)/\sqrt{2}$.
Hence, the determinant $\mathrm{det}\mathcal{T}=-\left(d+1\right)^{1-d^{2}}$
is \textit{negative}. In contrast, the determinant for Markovian channels
are always non-negative. Therefore, the partial corner transpose cannot
be obtained from Markovian channels. %
\footnote{In fact, for qubit channels, all rank-3 unital channels cannot even
be written as a product of two other channels (unless one of them
is a unitary channel). For these qubit exotic channels, an approach
based on convex decomposition of channels applies \cite{Sanders_PRL_2013}.
But for higher $d$ it is not known whether that will always work. %
}

We have obtained an explicit construction of $\left\{ U_{b^{\left(l\right)}}\right\} $
for the partial corner transpose channel with $d=3$, as detailed
in Appendix \ref{sec:Appendix: exotic}. For our channel construction
approach, the unitaries $U_{b^{\left(l\right)}}$ seem to be no more
difficult from other more conventional channels with the same rank. 

\textcolor{blue}{}%

\subsection{\textcolor{black}{Quantum Instrument and POVM}}

\textcolor{black}{The construction of CPTP maps can be further extended
if }\textit{\textcolor{black}{the intermediate measurement outcomes}}\textcolor{black}{{}
are part of the output together with the state of the quantum system,
which leads to an interesting class of quantum channel called a }\textit{\textcolor{black}{quantum
instrument }}\textcolor{black}{(QI) \cite{Wilde_book,Watrous_lec_notes,Blumoff_Chou_2016}.
QIs enable us to track both the classical measurement outcome and
the post-measurement state of the quantum system. Mathematically,
the quantum instrument has the following CPTP map: 
\begin{equation}
\rho\mapsto\mathcal{E}_{\mathrm{QI}}\left(\rho\right)=\sum_{\mu=1}^{M}\mathcal{E}_{\mu}(\rho)\otimes\ket{\mu}\bra{\mu},\label{eq:QI_map}
\end{equation}
where $\ket{\mu}\bra{\mu}$ are orthogonal projections of the measurement
device with $M$ classical outcomes, and $\mathcal{E}_{\mu}$ are
completely positive trace non-increasing maps, while $\sum_{\mu=1}^{M}\mathcal{E}_{\mu}(\rho)$
preserves the trace. Note that $\mathcal{E}_{\mu}(\rho)$ gives the
post-measurement state associated with outcome $\mu$. }

\begin{figure}
\centering{}\includegraphics{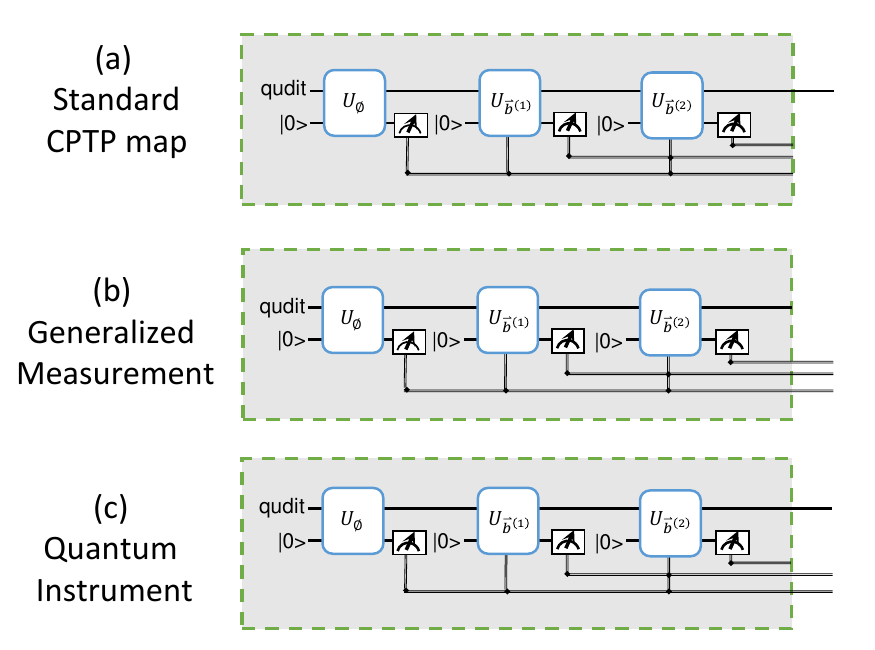}\protect\caption{(Color online) Three different types of CPTP maps. (a) To implement
a standard CPTP on the system qudit, all ancilla measurement records
should be thrown away; (b) A generalized measurement does not concern
the system state after measurement, so only the ancilla measurement
record is kept; (c) A quantum instrument keeps the both the post-measurement
state of the system and outcome $\mu$, encoded by the first $L_{1}$
bits of the ancilla measurement record. The remaining $L_{2}$ bits
of the measurement record are thrown away. In the figure, $L_{1}=2$
and $L_{2}=1$. \label{fig: QI-POVM-CPTP}}
\end{figure}

\textcolor{black}{As illustrated in Fig. \ref{fig: QI-POVM-CPTP},
our channel construction can implement the QI as follows. (1) Find
the minimum Kraus representation for $\mathcal{E}_{\mu}$ (each with
rank $J_{\mu}$) with Kraus operators $K_{\mu,j}$ for $j=1,\,2,\,\cdots,\, J_{\mu}$. (2)
Introduce binary labeling of these Kraus operators, $K_{\vec{b}^{(L)}}$,
where the binary label has length $L=L_{1}+L_{2}$ with the first
$L_{1}=\left\lceil \log_{2}M\right\rceil $ bits $b^{(L_{1})}$ to
encode $\mu$ and the remaining $L_{2}=\left\lceil \log_{2}\max_{\mu}(J^{\mu})\right\rceil $
bits to encode $j$ (padding with zero operators to make a total of
$2^{L}$ Kraus operators). (3) Use the quantum circuit with $L$ rounds
of adaptive evolution and ancilla measurement. (4) Output the final
state of the quantum system as well as $b^{(L_{1})}$ that encodes
$\mu$ associated with the $M$ possible classical outcomes. This
enables us to construct the arbitrary QI described in Eq. (\ref{eq:QI_map}).
The QI is a very useful tool for implementation of complicated conditional
evolution of the system. It can be used for quantum information processing
tasks that require measurement and adaptive control.}

If we remove the quantum system from the QI output, we effectively
implement a\textcolor{black}{{} positive operator valued measure (POVM),
which is also referred to as a generalized quantum measurement. A
POVM is a CPTP map from the quantum state of the system to the classical
state of the measurement device
\[
\rho\mapsto\mathcal{E}_{\mathrm{POVM}}\left(\rho\right)=\sum_{\mu=1}^{M}\mathrm{Tr}\left[\Pi_{\mu}\rho\right]\ket{\mu}\bra{\mu},
\]
which is characterized by a set of Hermitian positive semidefinite
operators $\{\Pi_{\mu}\}_{\mu=1}^{M}$ that sum to the identify operator
$\sum_{\mu}\Pi_{\mu}=\mathbb{I}$. For positive semidefinite $\Pi_{\mu}$,
we can decompose it as $\Pi_{\mu}=\sum_{j}K_{\mu,j}^{\dagger}K_{\mu,j}$
with a set of Kraus operators $\left\{ K_{\mu,j}\right\} _{j=1,\cdots,J_{\mu}}$.
Therefore, the circuit for the quantum instrument also implements
the POVM if we remove the quantum system from the QI output, $\mathcal{E}_{\mathrm{POVM}}\left(\rho\right)=\mathrm{Tr_{sys}}\left[\mathcal{E}_{\mathrm{QI}}\left(\rho\right)\right]$,
which reduces to the binary tree construction scheme of a POVM as
proposed by Andersson and Oi \cite{binary_tree_POVM}. A POVM can
be useful for quantum state discrimination. It is known to be impossible
for any detector to perfectly discriminate a set of non-orthogonal
quantum states. An optimal detector can achieve the so-called Hellstrom
bound \cite{Hellstrom_book_1976}, by properly designing a POVM (in
this case a PVM--projection valued measure). For example, in optical
communication, quadrature phase shift keying uses four coherent states
with different phases $\ket{\alpha}$, $\ket{i\alpha}$, $\ket{-\alpha}$
and $\ket{-i\alpha}$ to send two classical bits of information. With
our scheme it is straightforward to implement the optimal POVM given
in Ref. \cite{QPSK_PRA_1996}, which is a rank-4 POVM. }

\textcolor{black}{As summarized in Fig. \ref{fig: QI-POVM-CPTP},
we may classify three different situations for CPTP maps based on
the output: (a) standard quantum channel with the quantum system as
the output, (b) POVM with the classical measurement outcomes as the
output, (c) QI with both the quantum system and the classical measurement
outcomes for the output. In principle, all three situations can be
reduced to the standard quantum channel with an expanded quantum system
that includes an additional measurement device to keep track of the
classical measurement outcomes. In practice, however, it is much more
resource efficient to use a classical memory for classical measurement
outcomes, so that we can avoid working with the expanded quantum system.}

\section{Discussion \label{sec:Discussion}}

\textcolor{black}{So far, we have assumed a two-level ancilla for
our channel construction, which can be generalized to an ancilla with
higher dimensions. If we use an }\textit{\textcolor{black}{$s$}}\textcolor{black}{-dimensional
ancilla, we can use an s-ary tree construction of the quantum channel
with Kraus rank $N$, consisting of $\left\lceil \log_{s}N\right\rceil $
rounds of adaptive evolution and ancilla measurement. }

\textcolor{black}{We emphasize that the adaptive control is essential
for arbitrary channel construction with a small (low-dimensional)
ancilla. Without adaptive control, the constructed channel is a product
of channels, $\mathcal{T}=\cdots\mathcal{T}_{3}\mathcal{T}_{2}\mathcal{T}_{1}$,
and it excludes indivisible channels which cannot be constructed with
a single round of operation or decomposed into a product of non-unitary
channels \cite{divisibility}. Although the approach of Trotterization
and stroboscopic control can construct Markovian channels }\textit{\textcolor{black}{without}}\textcolor{black}{{}
adaptive control, that approach has an overhead that increases with
the duration of the Markovian evolution \cite{zanardi_lindbladian_sim},
while our construction has a bounded overhead that scales logarithmically
with the relevant dimensions of the quantum system.}

\textcolor{black}{Besides developing a control toolbox for quantum
information processing, our channel construction protocol may also
be useful for investigating open quantum systems, with the potential
advantages of reduced overhead in channel construction and the new
ingredient of indivisible channels, which are not accessible with
conventional reservoir engineering of Markovian channels \cite{Diehl_nat_phy_2008,Weimer_nat_phy_2010,Bardyn_topo_dissi_2013,Morigi_ion_chain_PRL,Diehl_chern_insu_PRA_2015}.}

\textcolor{black}{In experimental realizations, there will be imperfections
in the unitary gates $U_{b^{\left(l\right)}}$ and ancilla measurements.
Fortunately, the quantum circuit for channel construction only has
$n=\left\lceil \log_{2}\mbox{N}\right\rceil \le\left\lceil 2\log_{2}d\right\rceil $
rounds of gate and measurement. }If the error per round is $\epsilon$,
then the overall error rate of the channel construction is only $n\epsilon\sim\epsilon\log_{2}d$.
More rigorously, we may use the diamond norm distance $\epsilon_{\diamond}$
to upper bound the error associated with each round of operation \cite{Wilde_book},
and $n\epsilon_{\diamond}$ rigorously bounds the diamond norm distance
of the constructed quantum channel.

\section{Conclusion\label{sec:Conclusion}}

\textcolor{black}{We have provided an explicit procedure to construct
arbitrary CPTP maps, assisted by an ancilla qubit with QND readout
and adaptive control. Our construction has various applications, including system
initialization/stabilization, quantum error correction, Markovian
and exotic channel simulation, and generalized quantum measurement/quantum
instruments construction. Such a construction can be implemented with
circuit QED and various other physical platforms. }
\begin{acknowledgments}
\textcolor{black}{We thank Reinier Heeres, Phillip Reinhold, and Changling
Zou for helpful discussions. We acknowledge support from ARL-CDQI,
ARO (W911NF-14-1-0011, W911NF-14-1-0563), ARO MURI (W911NF-16-1-0349
), NSF (DMR-1609326, DGE-1122492), AFOSR MURI (FA9550-14-1- 0052,
FA9550-14-1-0015), the Alfred P. Sloan Foundation (BR2013-049), and
the Packard Foundation (2013-39273).}

\textit{\textcolor{black}{Note added}}\textcolor{black}{: While finalizing
the manuscript, the authors became aware of a related work on quantum
channels \cite{Christandl_CNOTS_2016}, which studies a different
way to construct a channel. In contrast to that work focusing on minimizing
the number of C-NOT gates, here we explicitly provide an efficient
protocol to construct quantum channels, propose a circuit QED implementation,
and} discuss various applications.%

\end{acknowledgments}

\bibliographystyle{apsrev4-1}
\bibliography{reference_list}

\begin{widetext}

\appendix

\section{Representations of Quantum Channels\label{Appendix: representation channels}}

In this appendix we review some basics on alternative ways a CPTP
map can be represented and how to convert back and forth between different
representations. Since our scheme favors the Kraus representation
as our ``canonical representation'', it is important to understand
how to convert a target channel in other representations to the Kraus
form.

\subsection{Superoperator Matrix Representation}

Since CPTP maps are linear in the density matrix $\rho$, we can treat
$\rho$ as a vector and write down the matrix form of the super-operator
$\mathcal{T}$, such that
\[
\tilde{\rho}_{ij}=\sum_{m,n}T_{ij,mn}\rho_{mn}
\]
or 
\[
\vec{\tilde{\rho}}=T\cdot\vec{\rho}
\]
 where $\tilde{\rho}=\mathcal{T}(\rho)$. This matrix form is particularly
useful when one considers the concatenation of channels. Applying
channel $\mathcal{T}_{1}$ first and then $\mathcal{T}_{2}$ results
in the overall channel represented by the matrix $T=T_{2}\cdot T_{1}$,
where ``$\cdot$'' indicates matrix multiplication. The matrix form
also allows one to characterize channels with the determinant, $\det(T)$.
One interesting property is that for Markovian channels or Kraus rank-2
channels, the determinant is always positive \cite{divisibility}.
The downside of this representation is that it is not obvious whether
a given $T$ qualifies as a CPTP map. We will need to convert it to
the Jamiolkowski/Choi matrix representation or Kraus representation
to verify that. Conversely, given a channel in Kraus form, the super-operator
matrix can be obtained straightforwardly,
\[
T=\sum_{i}^{N}K_{i}\otimes K_{i}^{*}.
\]

\subsection{Jamiolkowski/Choi Matrix Representation}

From the well known channel-state duality (Jamiolkowski-Choi isomorphism)
\cite{Jamiolkowski,Choi} we know that each channel $\mathcal{T}$
for a system with d-dimensional Hilbert space $\mathcal{H}$ corresponds
(one-to-one) to a state (a density matrix) on $\mathcal{H}\otimes\mathcal{H}$,
\[
\tau=(\mathcal{T}\otimes\mathcal{I})(\ket{\Omega}\bra{\Omega})
\]
 where $\ket{\Omega}=\frac{1}{\sqrt{d}}\sum_{i}\ket i\otimes\ket i$
is the maximally entangled state of the two subsystems. A closely
related matrix is the Choi matrix which is only a constant multiple
of the Jamiolkowski matrix, $M=d\,\tau$, where $d$ is the dimension
of the Hilbert space. A convenient fact to note is that $M$ and the
super-operator matrix $T$ are related in a simple way,

\[
T_{ij,mn}=M_{im,jn}.
\]
Being a density matrix, $\tau$ is Hermitian. Moreover $\tau$ is
semi-positive definite if and only if $\mathcal{T}$ is completely
positive; $\tau$ is normalized if $\mathcal{T}$ is trace preserving. 

It is straightforward to convert the Choi matrix $M$ to the Kraus
representation. If $M$ is diagonalized,
\[
M=\sum_{i}\lambda_{i}v_{i}v_{i}^{\dagger},
\]
where $v_{i}$ are $d^{2}$ dimensional eigenvectors of $\tau$, the
Kraus operators are obtained by rearranging $\sqrt{\lambda_{i}}v_{i}$
as $d\times d$ matrices. Clearly the number of non-zero eigenvalues
$\lambda_{i}$ is the Kraus rank of the corresponding channel. Later
we will often check the eigenvalue spectrum of the Choi matrix of
a channel to determine its Kraus rank. For numerical calculation we
usually make a truncation of the eigenvalues. For example, we may
set all eigenvalues smaller than $10^{-10}$ to 0.

\section{Proof of Quantum Channel Construction\label{sec:Appendix:-Proof of solution} }

We now prove that our channel construction correctly implements the
target CPTP map. To justify the channel construction, we need to show
that (a) the submatrices $\bra{b_{l+1}}U_{b^{\left(l\right)}}\ket 0$
fulfill the isometry condition 
\begin{equation}
\sum_{b_{l+1}=0,1}\left(\bra{b_{l+1}}U_{b^{\left(l\right)}}\ket 0\right)^{\dagger}\bra{b_{l+1}}U_{b^{\left(l\right)}}\ket 0=\mathbb{I}_{d\times d}\label{eq:cond1}
\end{equation}
 for all $b^{\left(l\right)}$ and $l=1,2,\cdots,L-1$, and (b) the
accumulated evolution along the binary tree indeed implements the
corresponding Kraus operator 
\begin{equation}
\left(\bra{b_{L}}U_{b^{\left(L-1\right)}}\ket 0\right)\cdots\left(\bra{b_{l+1}}U_{b^{\left(l\right)}}\ket 0\right)\cdots\left(\bra{b_{1}}U_{b^{\left(0\right)}}\ket 0\right)=K_{b^{\left(L\right)}}.\label{eq:cond2}
\end{equation}

First, we show that
\begin{equation}
V_{b^{\left(l\right)}}D_{b^{\left(l\right)}}^{2}V_{b^{\left(l\right)}}^{\dagger}=\sum_{b_{l+1},\cdots,b_{L}}K_{b^{\left(L\right)}}^{\dagger}K_{b^{\left(L\right)}}=\sum_{b_{l+1}}\left(\sum_{b_{l+2},\cdots,b_{L}}K_{b^{\left(L\right)}}^{\dagger}K_{b^{\left(L\right)}}\right)=\sum_{b_{l+1}=0,1}V_{b^{\left(l+1\right)}}D_{b^{\left(l+1\right)}}^{2}V_{b^{\left(l+1\right)}}^{\dagger}.\label{eq:tool_1}
\end{equation}
Since the right hand side is a sum of two non-negative matrices, we
also have the inequality 
\[
V_{b^{\left(l\right)}}D_{b^{\left(l\right)}}^{2}V_{b^{\left(l\right)}}^{\dagger}\ge V_{b^{\left(l+1\right)}}D_{b^{\left(l+1\right)}}^{2}V_{b^{\left(l+1\right)}}^{\dagger},
\]
which implies the same inequality for their support projections 
\[
V_{b^{\left(l\right)}}P_{b^{\left(l\right)}}V_{b^{\left(l\right)}}^{\dagger}\ge V_{b^{\left(l+1\right)}}P_{b^{\left(l+1\right)}}V_{b^{\left(l+1\right)}}^{\dagger}.
\]
Moreover, since $V_{b^{\left(l\right)}}V_{b^{\left(l\right)}}^{\dagger}=\mathbb{I}=V_{b^{\left(l+1\right)}}V_{b^{\left(l+1\right)}}^{\dagger}$
and $P_{b^{\left(l\right)}}^{\perp}=\mathbb{I}-P_{b^{\left(l\right)}}$,
we have
\begin{equation}
V_{b^{\left(l\right)}}P_{b^{\left(l\right)}}^{\perp}V_{b^{\left(l\right)}}^{\dagger}\le V_{b^{\left(l+1\right)}}P_{b^{\left(l+1\right)}}^{\perp}V_{b^{\left(l+1\right)}}^{\dagger},\label{eq:proj_containment}
\end{equation}
which demonstrates that the orthogonal support projection grows with
$l$. Using the fact that if projectors $P_{1}\le P_{2}$ then $P_{1}=P_{1}P_{2}P_{1}$,
we have
\[
V_{b^{\left(l\right)}}P_{b^{\left(l\right)}}^{\perp}V_{b^{\left(l\right)}}^{\dagger}=V_{b^{\left(l\right)}}P_{b^{\left(l\right)}}^{\perp}V_{b^{\left(l\right)}}^{\dagger}V_{b^{\left(l+1\right)}}P_{b^{\left(l+1\right)}}^{\perp}V_{b^{\left(l+1\right)}}^{\dagger}V_{b^{\left(l\right)}}P_{b^{\left(l\right)}}^{\perp}V_{b^{\left(l\right)}}^{\dagger},
\]
which is equivalent to
\begin{equation}
P_{b^{\left(l\right)}}^{\perp}=P_{b^{\left(l\right)}}^{\perp}V_{b^{\left(l\right)}}^{\dagger}V_{b^{\left(l+1\right)}}P_{b^{\left(l+1\right)}}^{\perp}V_{b^{\left(l+1\right)}}^{\dagger}V_{b^{\left(l\right)}}P_{b^{\left(l\right)}}^{\perp}.\label{eq:ProjRelation}
\end{equation}

Before we prove Eq. (\ref{eq:cond1}) and Eq. (\ref{eq:cond2}), we
first note that 
\begin{eqnarray*}
\bra{b_{l+1}}U_{b^{\left(l\right)}}\ket 0 & = & M_{b^{\left(l+1\right)}}M_{b^{\left(l\right)}}^{+}+\frac{1}{\sqrt{2}}Q_{b^{\left(l\right)}}\\
 & = & V_{b^{\left(l+1\right)}}D_{b^{\left(l+1\right)}}V_{b^{\left(l+1\right)}}^{\dagger}V_{b^{\left(l\right)}}D_{b^{\left(l\right)}}^{-1}V_{b^{\left(l\right)}}^{\dagger}+\frac{1}{\sqrt{2}}V_{b^{\left(l\right)}}P_{b^{\left(l\right)}}^{\perp}V_{b^{\left(l\right)}}^{\dagger}\\
 & = & V_{b^{\left(l+1\right)}}D_{b^{\left(l+1\right)}}V_{b^{\left(l+1\right)}}^{\dagger}V_{b^{\left(l\right)}}D_{b^{\left(l\right)}}^{-1}V_{b^{\left(l\right)}}^{\dagger}+\frac{1}{\sqrt{2}}V_{b^{\left(l+1\right)}}P_{b^{\left(l+1\right)}}^{\perp}V_{b^{\left(l+1\right)}}^{\dagger}V_{b^{\left(l\right)}}P_{b^{\left(l\right)}}^{\perp}V_{b^{\left(l\right)}}^{\dagger}\\
 & = & V_{b^{\left(l+1\right)}}\left[D_{b^{\left(l+1\right)}}V_{b^{\left(l+1\right)}}^{\dagger}V_{b^{\left(l\right)}}D_{b^{\left(l\right)}}^{-1}+\frac{1}{\sqrt{2}}P_{b^{\left(l+1\right)}}^{\perp}V_{b^{\left(l+1\right)}}^{\dagger}V_{b^{\left(l\right)}}P_{b^{\left(l\right)}}^{\perp}\right]V_{b^{\left(l\right)}}^{\dagger},
\end{eqnarray*}
where the third equality uses Eq. (\ref{eq:proj_containment}). Similarly,
\begin{eqnarray*}
\bra{b_{l+1}}U_{b^{\left(l\right)}}\ket 0 & = & K_{b^{\left(l+1\right)}}M_{b^{\left(l\right)}}^{+}+\frac{1}{\sqrt{2}}W_{b^{\left(l+1\right)}}V_{b^{\left(l+1\right)}}^{\dagger}Q_{b^{\left(l\right)}}\\
 & = & K_{b^{\left(l+1\right)}}V_{b^{\left(l\right)}}D_{b^{\left(l\right)}}^{-1}P_{b^{\left(l\right)}}+\frac{1}{\sqrt{2}}W_{b^{\left(l+1\right)}}V_{b^{\left(l+1\right)}}^{\dagger}V_{b^{\left(l\right)}}P_{b^{\left(l\right)}}^{\perp}V_{b^{\left(l\right)}}^{\dagger}\\
 & = & K_{b^{\left(l+1\right)}}V_{b^{\left(l\right)}}D_{b^{\left(l\right)}}^{-1}P_{b^{\left(l\right)}}+\frac{1}{\sqrt{2}}W_{b^{\left(l+1\right)}}V_{b^{\left(l+1\right)}}^{\dagger}V_{b^{\left(l+1\right)}}P_{b^{(l+1)}}^{\perp}V_{b^{\left(l+1\right)}}^{\dagger}V_{b^{\left(l\right)}}P_{b^{\left(l\right)}}^{\perp}V_{b^{\left(l\right)}}^{\dagger}\\
 & = & \left(K_{b^{\left(l+1\right)}}V_{b^{\left(l\right)}}D_{b^{\left(l\right)}}^{-1}P_{b^{\left(l\right)}}+\frac{1}{\sqrt{2}}W_{b^{\left(l+1\right)}}P_{b^{\left(l+1\right)}}^{\perp}V_{b^{\left(l+1\right)}}^{\dagger}V_{b^{\left(l\right)}}P_{b^{\left(l\right)}}^{\perp}\right)V_{b^{\left(l\right)}}^{\dagger}.
\end{eqnarray*}

To prove Eq. (\ref{eq:cond1}) for $l=0,1,\cdots,L-2$, we use 
\begin{align*}
 & \sum_{b_{l+1}=0,1}\left(\bra{b_{l+1}}U_{b^{\left(l\right)}}\ket 0\right)^{\dagger}\bra{b_{l+1}}U_{b^{\left(l\right)}}\ket 0\\
= & V_{b^{\left(l\right)}}\left[\sum_{b_{l+1}=0,1}\left(D_{b^{\left(l+1\right)}}V_{b^{\left(l+1\right)}}^{\dagger}V_{b^{\left(l\right)}}D_{b^{\left(l\right)}}^{-1}P_{b^{\left(l\right)}}\right)^{\dagger}D_{b^{\left(l+1\right)}}V_{b^{\left(l+1\right)}}^{\dagger}V_{b^{\left(l\right)}}D_{b^{\left(l\right)}}^{-1}P_{b^{\left(l\right)}}+\frac{1}{2}\sum_{b_{l+1}=0,1}\left(P_{b^{\left(l+1\right)}}^{\perp}V_{b^{\left(l+1\right)}}^{\dagger}V_{b^{\left(l\right)}}P_{b^{\left(l\right)}}^{\perp}\right)^{\dagger}P_{b^{\left(l+1\right)}}^{\perp}V_{b^{\left(l+1\right)}}^{\dagger}V_{b^{\left(l\right)}}P_{b^{\left(l\right)}}^{\perp}\right]V_{b^{\left(l\right)}}^{\dagger}\\
= & V_{b^{\left(l\right)}}\left[P_{b^{\left(l\right)}}D_{b^{\left(l\right)}}^{-1}V_{b^{\left(l\right)}}^{\dagger}\left(\sum_{b_{l+1}=0,1}V_{b^{\left(l+1\right)}}D_{b^{\left(l+1\right)}}^{2}V_{b^{\left(l+1\right)}}^{\dagger}\right)V_{b^{\left(l\right)}}D_{b^{\left(l\right)}}^{-1}P_{b^{\left(l\right)}}+\frac{1}{2}\sum_{b_{l+1}=0,1}P_{b^{\left(l\right)}}^{\perp}V_{b^{\left(l\right)}}^{\dagger}V_{b^{\left(l+1\right)}}P_{b^{\left(l+1\right)}}^{\perp}V_{b^{\left(l+1\right)}}^{\dagger}V_{b^{\left(l\right)}}P_{b^{\left(l\right)}}^{\perp}\right]V_{b^{\left(l\right)}}^{\dagger}\\
= & V_{b^{\left(l\right)}}\left[P_{b^{\left(l\right)}}+P_{b^{\left(l\right)}}^{\perp}\right]V_{b^{\left(l\right)}}^{\dagger}\\
= & V_{b^{\left(l\right)}}IV_{b^{\left(l\right)}}^{\dagger}\\
= & I
\end{align*}
where the first equality uses the orthogonality property $P_{b^{\left(l\right)}}P_{b^{\left(l\right)}}^{\perp}=0$,
the third equality uses Eq. (\ref{eq:tool_1}) %
{} and Eq. (\ref{eq:ProjRelation}). Similarly, we can prove Eq. (\ref{eq:cond1})
for $l=L-1$. 

To prove Eq. (\ref{eq:cond2}), we have
\begin{align*}
 & \bra{b_{L}}U_{b^{\left(L-1\right)}}\ket 0\cdots\bra{b_{2}}U_{b^{\left(1\right)}}\ket 0\bra{b_{1}}U_{b^{\left(0\right)}}\ket 0\\
= & \left(K_{b^{\left(L\right)}}V_{b^{\left(L-1\right)}}D_{b^{\left(L-1\right)}}^{-1}P_{b^{\left(L-1\right)}}V_{b^{\left(L-1\right)}}^{\dagger}\right)\cdots\left(V_{b^{\left(l+1\right)}}D_{b^{\left(l+1\right)}}V_{b^{\left(l+1\right)}}^{\dagger}V_{b^{\left(l\right)}}D_{b^{\left(l\right)}}^{-1}P_{b^{\left(l\right)}}V_{b^{\left(l\right)}}^{\dagger}\right)\cdots\left(V_{b^{\left(2\right)}}D_{b^{\left(1\right)}}V_{b^{\left(1\right)}}^{\dagger}\right)\\
= & K_{b^{\left(L\right)}}\left(V_{b^{\left(L-1\right)}}P_{b^{\left(L-1\right)}}V_{b^{\left(L-1\right)}}^{\dagger}\right)\cdots\left(V_{b^{\left(l\right)}}P_{b^{\left(l\right)}}V_{b^{\left(l\right)}}^{\dagger}\right)\cdots\left(V_{b^{\left(1\right)}}D_{b^{\left(1\right)}}V_{b^{\left(1\right)}}^{\dagger}\right)\\
= & K_{b^{\left(L\right)}}\left(V_{b^{\left(L-1\right)}}P_{b^{\left(L-1\right)}}V_{b^{\left(L-1\right)}}^{\dagger}\right)\\
= & \left(W_{b^{\left(L\right)}}D_{b^{\left(L\right)}}V_{b^{\left(L\right)}}^{\dagger}\right)\left(V_{b^{\left(L-1\right)}}P_{b^{\left(L-1\right)}}V_{b^{\left(L-1\right)}}^{\dagger}\right)\\
= & \left(W_{b^{(L)}}D_{b^{(L)}}P_{b^{(L)}}V_{b^{(L)}}^{\dagger}\right)\left(V_{b^{\left(L-1\right)}}P_{b^{\left(L-1\right)}}V_{b^{\left(L-1\right)}}^{\dagger}\right)\\
= & \left(W_{b^{(L)}}D_{b^{(L)}}V_{b^{(L)}}^{\dagger}V_{b^{(L)}}P_{b^{(L)}}V_{b^{(L)}}^{\dagger}\right)\left(V_{b^{\left(L-1\right)}}P_{b^{\left(L-1\right)}}V_{b^{\left(L-1\right)}}^{\dagger}\right)\\
= & W_{b^{(L)}}D_{b^{(L)}}V_{b^{(L)}}^{\dagger}V_{b^{(L)}}P_{b^{(L)}}V_{b^{(L)}}^{\dagger}\\
= & W_{b^{(L)}}D_{b^{(L)}}P_{b^{(L)}}V_{b^{(L)}}^{\dagger}\\
= & K_{b^{\left(L\right)}},
\end{align*}
where the first equality only has one non-zero product, because all
other terms vanish due to the orthogonality property $P_{b^{\left(l\right)}}P_{b^{\left(l\right)}}^{\perp}=0$
and $P_{b^{\left(0\right)}}^{\perp}=0$, the second equality exploits
$V_{b^{\left(l\right)}}^{\dagger}V_{b^{\left(l\right)}}=I$, $D_{b^{\left(l\right)}}^{-1}P_{b^{\left(l\right)}}D_{b^{\left(l\right)}}=P_{b^{\left(l\right)}}$
and $V_{b^{\left(0\right)}}=D_{b^{\left(0\right)}}=P_{b^{\left(0\right)}}=\mathbb{I}$,
and the third and the last but two equalities require the projection
relation $\left(V_{b^{\left(l\right)}}P_{b^{\left(l\right)}}V_{b^{\left(l\right)}}^{\dagger}\right)\left(V_{b^{\left(l-1\right)}}P_{b^{\left(l-1\right)}}V_{b^{\left(l-1\right)}}^{\dagger}\right)=\left(V_{b^{\left(l\right)}}P_{b^{\left(l\right)}}V_{b^{\left(l\right)}}^{\dagger}\right)$. 

Therefore, we have proven both Eq. (\ref{eq:cond1}) and Eq. (\ref{eq:cond2}),
which fully justify our explicit construction of the CPTP map.

\section{Explicit Circuits for an Example Exotic channel\label{sec:Appendix: exotic}}

We show an explicit construction of the isometries needed for the
construction of the exotic channel 
\[
\mathcal{T}(\rho)=\frac{\rho^{T_{c}}+\mathbb{I}\,\mathrm{Tr}(\rho)}{1+d}
\]
 for the case of $d=3$. %
{} 
\[
\left(\begin{array}{c}
\bra 0U_{b^{(0)}}\ket 0\\
\bra 1U_{b^{(0)}}\ket 0
\end{array}\right)=\left(\begin{array}{ccc}
\frac{\sqrt{10+\sqrt{2}}}{4}\\
 & \frac{\sqrt{2+\frac{1}{\sqrt{2}}}}{2}\\
 &  & \frac{\sqrt{10+\sqrt{2}}}{4}\\
\frac{\sqrt{6-\sqrt{2}}}{4}\\
 & \frac{\sqrt{2-\frac{1}{\sqrt{2}}}}{2}\\
 &  & \frac{\sqrt{6-\sqrt{2}}}{4}
\end{array}\right),
\]

\[
\left(\begin{array}{c}
\bra 0U_{b^{(1)}=0}\ket 0\\
\bra 1U_{b^{(1)}=0}\ket 0
\end{array}\right)=\left(\begin{array}{ccc}
\frac{\sqrt{29+2\sqrt{2}}}{7}\\
 & \sqrt{(3+\sqrt{2})/7}\\
 &  & \frac{\sqrt{29+2\sqrt{2}}}{7}\\
\frac{2}{\sqrt{10+\sqrt{2}}}\\
 & \frac{1}{\sqrt{2+\frac{1}{\sqrt{2}}}}\\
 &  & \frac{2}{\sqrt{10+\sqrt{2}}}
\end{array}\right),\,\left(\begin{array}{c}
\bra 0U_{b^{(1)}=1}\ket 0\\
\bra 1U_{b^{(1)}=1}\ket 0
\end{array}\right)=\left(\begin{array}{ccc}
\sqrt{2(6+\sqrt{2})/17}\\
 & 0\\
 &  & \sqrt{2(6+\sqrt{2})/17}\\
\sqrt{(5-2\sqrt{2})/17}\\
 & 1\\
 &  & \sqrt{(5-2\sqrt{2})/17}
\end{array}\right),\ 
\]

\[
\left(\begin{array}{c}
\bra 0U_{b^{(2)}=00}\ket 0\\
\bra 1U_{b^{(2)}=00}\ket 0
\end{array}\right)=\left(\begin{array}{ccc}
\sqrt{(5+2\sqrt{2})/17}\\
 & 1\\
 &  & \sqrt{(5+2\sqrt{2})/17}\\
-\frac{2}{\sqrt{6+\sqrt{2}}}\\
 & 0\\
 &  & \frac{2}{\sqrt{6+\sqrt{2}}}
\end{array}\right),\,\left(\begin{array}{c}
\bra 0U_{b^{(2)}=01}\ket 0\\
\bra 1U_{b^{(2)}=01}\ket 0
\end{array}\right)=\left(\begin{array}{ccc}
0 & 0 & 1\\
0 & 0 & 0\\
1 & 0 & 0\\
0 & 0 & 0\\
0 & 0 & 0\\
0 & 1 & 0
\end{array}\right),
\]

\[
\left(\begin{array}{c}
\bra 0U_{b^{(2)}=10}\ket 0\\
\bra 1U_{b^{(2)}=10}\ket 0
\end{array}\right)=\left(\begin{array}{ccc}
0 & 1 & 0\\
0 & 0 & 1\\
0 & 0 & 0\\
0 & 0 & 0\\
1 & 0 & 0\\
0 & 0 & 0
\end{array}\right),\ \left(\begin{array}{c}
\bra 0U_{b^{(2)}=11}\ket 0\\
\bra 1U_{b^{(2)}=11}\ket 0
\end{array}\right)=\left(\begin{array}{ccc}
0 & \sqrt{(4+\sqrt{2})/7} & 0\\
0 & 0 & 0\\
0 & 0 & 0\\
1\\
 & -\sqrt{(3-\sqrt{2})/7}\\
 &  & 1
\end{array}\right).
\]

\end{widetext}
\end{document}